\documentclass[12pt]{article}
\usepackage[T1]{fontenc}
\usepackage{aer}
\usepackage{amsmath,amssymb,array,booktabs,float,graphicx,natbib,setspace,tabularx,tikz}
\usetikzlibrary{arrows.meta,backgrounds,calc,fit,positioning,shapes.geometric}
\usepackage[margin=1in]{geometry}
\usepackage[hidelinks]{hyperref}
\usepackage{microtype}

\hypersetup{
  pdftitle={Tabular Foundation Models and the Unity of Economic Behaviour},
  pdfauthor={Victor H. Aguiar},
  pdfsubject={Tabular foundation models as unified predictive models of economic behaviour},
  pdfkeywords={random utility, bounded rationality, discrete choice, foundation models, joint prediction}
}

\newcommand{\TabFM}{TabFM}

\title{Tabular Foundation Models and the Unity of Economic Behaviour}
\author{Victor H. Aguiar\thanks{Department of Economics, Simon Fraser
University, 8888 University Drive, Burnaby, British Columbia V5A 1S6, Canada.
Email: \href{mailto:vaguiarl@sfu.ca}{vaguiarl@sfu.ca}. I thank the Digital
Research Alliance of Canada for computational resources. This paper draws on
computational work supported by the Alliance and the Government of Canada.}}
\date{August 2026}

\begin{document}
\maketitle

\begin{abstract}
\noindent Economics uses different behavioural models for risk, time, losses,
valuation, and social choice. I study a unified choice experiment in which the
same decision makers face all these domains. I hide a decision maker's choices
in one domain and ask a frozen tabular foundation model to recover them from
that decision maker's choices elsewhere and labelled choices by other
participants. The foundation model improves on the training-sample median, and
the gain disappears when visible choices are shuffled across decision makers. I then
estimate one random-utility model over the foundation model's learned
representation. This structural model applies the same utility function in
every domain, retains most of the foundation model's reduction in prediction
error, predicts domains excluded from utility estimation, and reproduces how
behavioural measures co-move across people. The resulting model separates
three objects: a learned common choice domain, one systematic utility function
on that domain, and one random component that generates stochastic choice on
observed menus.

\end{abstract}

\noindent\textit{Journal of Economic Literature (JEL) codes:} C45, C53, D81, D90.

\setstretch{1.40}

\section{Introduction}

A central ambition of economics is to explain different decision domains with
a single unified model. Random utility already supplies a formally universal
choice principle. Its abstract domain may contain any alternative: a decision
maker assigns utility to the available alternatives and chooses the one with
the highest utility after a nonsystematic shock is added
\citep{McFadden1974,GulPesendorfer2006,Train2009}. But this formal generality
does not say how unlike alternatives should be represented. A lottery arrives
as probabilities and prizes, a delayed reward as a date and a payment, and an
allocation as payoffs to different people. Standard applications therefore
build separate representations and utility specifications for risk, time,
losses, valuation, and social choice. Random utility is unified at the level
of the choice rule but silent at the level of representation.

This paper separates the representation of choice from its valuation. The new
model has three components. First, a frozen foundation encoder maps the
decision problem and behavioural context into a rich state; combining that
state with each feasible response plan constructs a common choice domain.
Second, one systematic utility function values every feasible point in that
domain. Third, one common shock law adds nonsystematic utility and turns those
valuations into stochastic choice. The raw problem changes across risk, time,
and social domains; the learned choice domain, systematic valuation map, and
shock law are common.

Figure~\ref{fig:unifying-architecture} states the architecture. Encoding
constructs the choice domain; systematic utility values it; random shocks
generate choice. The second and third components form a random utility model.
Systematic utility carries the ranking over observed response plans; extending
welfare comparisons to new menus additionally requires the learned
representation of primitive actions to remain stable. This separation
preserves the traditional role of utility while allowing the domain on which
it operates to be learned from broad behavioural data.

\definecolor{DomainBlue}{HTML}{2A6592}
\definecolor{DomainTeal}{HTML}{21867A}
\definecolor{UtilityGold}{HTML}{C58B24}
\definecolor{ChoiceRed}{HTML}{A94B4B}
\definecolor{Ink}{HTML}{26313A}
\definecolor{SoftGrey}{HTML}{F4F5F6}

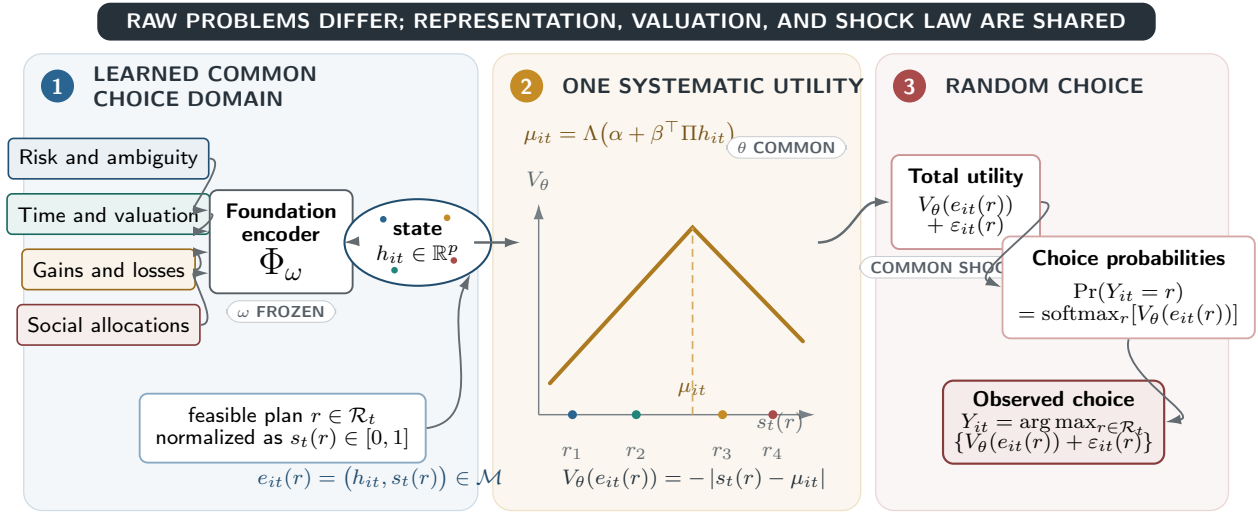
\begin{figure}[!t]
\centering
\resizebox{\textwidth}{!}{%
\begin{tikzpicture}[
  x=1cm,
  y=1cm,
  font=\sffamily\scriptsize,
  >={Latex[length=2.1mm,width=1.35mm]},
  flow/.style={-{Latex[length=2.1mm,width=1.35mm]},draw=Ink!72,line width=0.72pt},
  domain/.style={rounded corners=2pt,draw=#1!78!black,fill=#1!10,
    line width=0.62pt,minimum width=2.05cm,minimum height=0.54cm,align=center},
  box/.style={rounded corners=3pt,draw=Ink!78,fill=white,line width=0.72pt,
    align=center,inner xsep=6pt,inner ysep=5pt},
  tag/.style={rounded corners=5pt,fill=white,draw=Ink!24,inner xsep=4pt,
    inner ysep=2pt,font=\sffamily\tiny\bfseries,text=Ink!78},
  header/.style={font=\sffamily\scriptsize\bfseries,text=Ink,align=left},
  formula/.style={font=\scriptsize,align=center,text=Ink}
]

\begin{scope}[on background layer]
  \path[rounded corners=7pt,fill=DomainBlue!5,draw=DomainBlue!22,line width=0.5pt]
    (0,0.20) rectangle (6.05,6.30);
  \path[rounded corners=7pt,fill=UtilityGold!7,draw=UtilityGold!25,line width=0.5pt]
    (6.25,0.20) rectangle (11.15,6.30);
  \path[rounded corners=7pt,fill=ChoiceRed!5,draw=ChoiceRed!22,line width=0.5pt]
    (11.35,0.20) rectangle (16.05,6.30);
\end{scope}

\node[circle,fill=DomainBlue,text=white,minimum size=0.44cm,
  inner sep=0pt,font=\sffamily\scriptsize\bfseries] at (0.45,5.87) {1};
\node[header,anchor=west] at (0.78,5.87) {LEARNED COMMON\\CHOICE DOMAIN};
\node[circle,fill=UtilityGold,text=white,minimum size=0.44cm,
  inner sep=0pt,font=\sffamily\scriptsize\bfseries] at (6.70,5.87) {2};
\node[header,anchor=west] at (7.03,5.87) {ONE SYSTEMATIC UTILITY};
\node[circle,fill=ChoiceRed,text=white,minimum size=0.44cm,
  inner sep=0pt,font=\sffamily\scriptsize\bfseries] at (11.80,5.87) {3};
\node[header,anchor=west] at (12.13,5.87) {RANDOM CHOICE};

\node[domain=DomainBlue] (risk) at (1.13,4.90) {Risk and ambiguity};
\node[domain=DomainTeal] (time) at (1.13,4.16) {Time and valuation};
\node[domain=UtilityGold] (loss) at (1.13,3.42) {Gains and losses};
\node[domain=ChoiceRed] (social) at (1.13,2.68) {Social allocations};

\node[box,minimum width=1.65cm,minimum height=1.18cm] (encoder) at (3.43,3.79)
  {\textbf{Foundation}\\[-1pt]\textbf{encoder}\\[2pt]
   {\large $\Phi_{\omega}$}};
\node[tag,anchor=north] at ($(encoder.south)+(0,-0.08)$) {$\omega$ FROZEN};

\draw[flow] (risk.east) to[out=0,in=180] ($(encoder.west)+(0,0.42)$);
\draw[flow] (time.east) to[out=0,in=180] ($(encoder.west)+(0,0.14)$);
\draw[flow] (loss.east) to[out=0,in=180] ($(encoder.west)+(0,-0.14)$);
\draw[flow] (social.east) to[out=0,in=180] ($(encoder.west)+(0,-0.42)$);

\node[ellipse,draw=DomainBlue!75!black,fill=white,line width=0.78pt,
  minimum width=1.36cm,minimum height=1.14cm,align=center] (state) at (5.25,3.78)
  {\textbf{state}\\$h_{it}\in\mathbb R^p$};
\foreach \dx/\dy/\cc in {-0.46/0.31/DomainBlue,-0.31/-0.37/DomainTeal,
  0.39/0.33/UtilityGold,0.48/-0.24/ChoiceRed}{
  \fill[\cc] ($(state.center)+(\dx,\dy)$) circle (1.25pt);
}
\draw[flow] (encoder.east) -- (state.west);

\node[box,draw=DomainBlue!55,minimum width=2.12cm,minimum height=0.62cm]
  (plan) at (3.45,1.30) {feasible plan $r\in\mathcal R_t$\\[-1pt]
  normalized as $s_t(r)\in[0,1]$};
\draw[flow] (plan.east) to[out=8,in=245] (state.south east);
\node[formula,text=DomainBlue!75!black] at (4.75,0.63)
  {$e_{it}(r)=\bigl(h_{it},s_t(r)\bigr)\in\mathcal M$};

\draw[flow,line width=0.95pt] (6.00,3.78) -- (6.62,3.78);
\node[formula,anchor=south,text=UtilityGold!60!black] at (8.08,4.91)
  {$\mu_{it}=\Lambda\!\left(\alpha+\beta^{\top}\Pi h_{it}\right)$};
\node[tag] at (10.16,5.05) {$\theta$ COMMON};

\begin{scope}[shift={(6.86,1.48)}]
  \draw[->,draw=Ink!62,line width=0.55pt] (0,0) -- (3.68,0)
    node[below=3pt,anchor=east,text=Ink!75] {$s_t(r)$};
  \draw[->,draw=Ink!62,line width=0.55pt] (0,0) -- (0,2.82)
    node[above=1pt,anchor=south,text=Ink!75] {$V_{\theta}$};
  \draw[UtilityGold!88!black,line width=1.55pt,line cap=round]
    (0.15,0.42) -- (2.05,2.50) -- (3.52,0.98);
  \draw[densely dashed,draw=UtilityGold!72,line width=0.65pt]
    (2.05,0) -- (2.05,2.50);
  \node[above=3pt,text=UtilityGold!65!black] at (2.05,0) {$\mu_{it}$};
  \foreach \xx/\lab/\cc in {0.45/$r_1$/DomainBlue,1.30/$r_2$/DomainTeal,
    2.45/$r_3$/UtilityGold,3.12/$r_4$/ChoiceRed}{
    \fill[\cc] (\xx,0) circle (1.7pt);
    \node[below=8pt,text=Ink!72] at (\xx,0) {\lab};
  }
\end{scope}
\node[formula,text=Ink] at (8.93,0.63)
  {$V_{\theta}\!\left(e_{it}(r)\right)=-\left|s_t(r)-\mu_{it}\right|$};

\node[box,draw=ChoiceRed!60,minimum width=1.90cm,minimum height=0.84cm]
  (shock) at (12.55,4.32) {\textbf{Total utility}\\[1pt]
  $V_{\theta}(e_{it}(r))$\\[-2pt]$+\;\varepsilon_{it}(r)$};
\node[tag,anchor=north] at ($(shock.south)+(0,-0.08)$) {COMMON SHOCK LAW};

\node[box,draw=ChoiceRed!48,minimum width=2.05cm,minimum height=1.18cm]
  (prob) at (14.72,3.18) {\textbf{Choice probabilities}\\[3pt]
  $\Pr(Y_{it}=r)$\\[-1pt]
  $=\operatorname{softmax}_{r}\!\left[V_{\theta}(e_{it}(r))\right]$};
\draw[flow,line width=0.95pt] (10.58,3.78) to[out=0,in=180] (shock.west);
\draw[flow] (shock.east) to[out=0,in=155] (prob.west);

\node[rounded corners=4pt,draw=ChoiceRed!80!black,fill=ChoiceRed!11,
  line width=0.82pt,minimum width=2.25cm,minimum height=0.72cm,align=center]
  (choice) at (13.72,1.38) {\textbf{Observed choice}\\[-1pt]
  $Y_{it}=\arg\max_{r\in\mathcal R_t}$\\[-2pt]
  $\{V_{\theta}(e_{it}(r))+\varepsilon_{it}(r)\}$};
\draw[flow] (prob.south) to[out=255,in=10] (choice.east);

\node[rounded corners=5pt,fill=Ink,text=white,inner xsep=11pt,inner ysep=4pt,
  font=\sffamily\scriptsize\bfseries,anchor=south] at (8.03,6.47)
  {RAW PROBLEMS DIFFER; REPRESENTATION, VALUATION, AND SHOCK LAW ARE SHARED};

\end{tikzpicture}%
}
\caption{A unified random-utility model in a learned choice domain}
\label{fig:unifying-architecture}
\begin{minipage}{0.96\textwidth}
\footnotesize
\textit{Notes:} The frozen encoder maps heterogeneous decision problems and
behavioural context into a common state. Combining that state with each
feasible response plan produces an object in the learned domain $\mathcal M$.
One estimated systematic utility function ranks those objects in every domain;
one common shock law generates stochastic choice on the observed menu. The
representation preserves domain-specific context, while valuation and choice
remain economically unified.
\end{minipage}
\end{figure}

Standard applications of random utility choose a domain-specific
representation and a valuation rule together. The hybrid learns a common
representation and disciplines its map into utility. Ordered
random utility locates the same restriction in a simpler setting: empirical
content lies in the map from types to utilities, not in the type distribution
alone
\citep{ApesteguiaBallesterLu2017,ApesteguiaBallester2025}. The learned state is
richer than an ordered type, but one map still carries it into utility. In an
action-level extension, risk aversion, present bias, loss aversion, and
inequality aversion would appear as economically meaningful restrictions on
the composite map from primitives to utility, rather than as unrelated
prediction heads.

The empirical literature measures several preferences in the same people
\citep{BarskyEtAl1997,ChoiEtAl2014,FalkEtAl2018}. \citet{ChapmanEtAl2023}
provide the experimental setting needed here. The same participants make
incentivized decisions spanning the authors' six broad preference domains. For
prediction, I partition the 26 raw multiple-price-list screens underlying 12
of their measures into eight operational domains. Those decisions share
structure, but no model they examine spans the collection. I impose a stronger
test: hide one domain, predict it from choices elsewhere, and ask whether one
valuation rule can make every prediction. Reassigning the visible choices
across decision makers preserves every task marginal but destroys the coherent
behavioural history.

The implementation uses Google's Tabular Foundation Model (TabFM), pretrained
on synthetic tables and frozen before this analysis \citep{KongDas2026}. The
data contain 26 multiple price list screens completed by 1,000 adults. In the
held out test, the model reduces mean absolute switch row error by 17 percent
relative to the training population median. Reassigning behavioural histories
across decision makers eliminates the gain. In participant cross fits, it also
outperforms Extra Trees and tuned CatBoost given identical features and context
observations. Choices contain nearly all of the useful predictive information.
Frozen does not mean label free: 100 target task outcomes enter as context, and
the model infers their relation to the query without changing its weights.

The main result places random utility on the foundation representation. A fixed
projection and one coefficient vector map each learned state into a choice
index. Each feasible switch plan receives systematic utility according to its
distance from that index; type I extreme value shocks give conditional logit
probabilities. One index and one utility rule serve all 26 tasks, but their
coefficients are estimated from the human training sample.

This severe restriction preserves most of the predictive content. The common
utility index places 65.83 percent of choices within two rows, compared with
40.42 percent for the task population median and 71.04 percent for TabFM's
original nonlinear decoder. Measured by normalized error, it retains 85
percent of the foundation model's improvement over the population forecast.
When every outcome from the target domain is excluded while fitting the
utility index, accuracy remains 63.16 percent and the hybrid retains 78 percent
of the original improvement. The representation still receives the same
target task context as the frozen model; the excluded domain test isolates
transfer of the valuation rule. One valuation direction learned outside a
domain therefore predicts decisions inside it. This establishes transfer of an
estimated utility map, not zero shot learning of utility.

The hybrid also reproduces the ordering of the central empirical pattern in
Econographics.
Across the 66 pairwise relationships among 12 author-defined behavioural
constructs, the common utility model's measurement-error-corrected correlation
matrix has rank correlation 0.939 with the observed matrix. Estimating the
valuation rule without any target-domain outcomes leaves the alignment at
0.935. Thus TabFM supplies a predictive unified model for the Chapman battery,
and one classical random utility map recovers both its predictions and the
joint organization of behaviour. It overstates the average strength of
dependence, so the result is a common organization of the correlation geometry,
not a complete independent-shock explanation of its level.

The internal geometry supports this interpretation. Same domain tasks remain
more similar across decision makers than different domain tasks, and the gap
grows with depth, although all domains use almost the same block update
schedule. A replication in the public Self Regulation Ontology reaches the
same operational conclusion across 37 behavioural tasks
\citep{EisenbergEtAl2019}.

The contribution has two levels. First, one estimated utility map unifies
valuation across domains. Second, the results identify the remaining step
toward a universal structural model. TabFM adapts through context with fixed
weights; the utility layer becomes fixed only after sample estimation. Matching
that capability requires a common economic learning rule that infers explicit
utility from new context without refitting its own parameters.

\section{Prediction Problem and Data}
\label{sec:prediction-data}

\subsection{A test of unified prediction}

Index participants by $i=1,\ldots,N$ and MPL tasks by $t=1,\ldots,T$. Let
$d(t)$ denote the economic domain of task $t$. The outcome $Y_{it}$ is the first
row on which participant $i$ selects the right-hand option. Because the
interface enforces a single switch, this code identifies the complete binary
choice vector within a task. Some screens permit never switching, which is a
separate feasible response.

Let $X_i$ contain nonchoice information and let $Z_t$ describe the target
mechanism, payoff ladder, and feasible response set. When domain $d$ is hidden,
the object of interest is
\begin{equation}
 P\!\left(Y_{i,d}\mid X_i,Y_{i,-d},Z_d,\mathcal D\right),
 \label{eq:joint-object}
\end{equation}
where $Y_{i,d}$ is the vector of hidden tasks, $Y_{i,-d}$ is the visible choice
history, and $\mathcal D$ is a labelled context containing other participants.
Point forecasts summarize the centre of each coordinate. Marginal predictive
distributions describe uncertainty in each task. Their dependence describes
whether the forecast preserves the fact that all hidden choices belong to one
decision maker.

A predictor is unified in the sense tested here when the prediction operator,
feature rules, and decoder remain common across domains and the visible choices
of the same decision maker improve forecasts of the hidden domain. Section
\ref{sec:embedded-utility} strengthens this operational definition by imposing
one choice index on the common representation. Its action-level extension maps
the feasible switch plans into random utility; a stronger action-level form
maps lotteries, dates, and social payoffs into the same valuation rule
\citep{McFadden1974,Train2009}.

Let $A_{it}\in[0,1]$ be the share of feasible rows on which option A is chosen,
and let $\widehat A_{it}^{m}$ be the forecast from model $m$. The primary
normalized participant loss is
\begin{equation}
 L_i^m=\frac{1}{|\mathcal T_i|}\sum_{t\in\mathcal T_i}
 \left|A_{it}-\widehat A_{it}^{m}\right|,
 \qquad F_i^m=1-L_i^m,
 \label{eq:participant-fitness}
\end{equation}
where $\mathcal T_i$ is the set of masked tasks. I call $F_i^m$ behavioural
prediction fitness. It lies between zero and one and gives every participant
equal weight. Raw switch-row mean absolute error (MAE), exact accuracy, and the
share within one or two rows are reported alongside it.

The main test has three parts. First, the frozen model must outperform a target-
and-context-matched population forecast. Second, that advantage must deteriorate
when every visible task retains its marginal distribution but the task values
are independently reassigned across participants. Third, the same frozen checkpoint
must serve every target. The first comparison establishes useful conditional
information. The second identifies coherent within-decision-maker behaviour as its
source. The third rules out a collection of separately estimated behavioural
models disguised as one procedure.

A locally trained nonlinear learner receives the same features and labels and
estimates a new mapping for each target. It measures how much predictive
content is already present in the joint choice vector. The frozen foundation
model carries one parameter vector, learned before Econographics, across all
targets. Their comparison separates the value of local information from the
value of a portable inference rule. Fixed-parameter portability addresses one
concern raised by the Lucas critique \citep{Lucas1976}. It is not by itself
policy invariance: the learned representation must also remain stable when
payoffs and information change.

\subsection{Econographics}

\citet{ChapmanEtAl2023} recruited a matched and survey-weighted YouGov sample of
1,000 United States adults. Their study elicits 21 behavioural measures and uses
repeated elicitations, obviously related instrumental variables (ORIV), and principal
components to study their joint structure \citep{GillenSnowbergYariv2019}. I
use their public replication deposit \citep{ChapmanEtAl2022Data}, but study a
different outcome: prediction of the raw switch responses from which 12 of
their measures are constructed.

The panel contains 26 MPL screens covering eight analysis domains: known risk,
ambiguity, compound lotteries, probability weighting, time discounting,
valuation, losses, and distributional preferences. The screens include risk
choices with certain and lottery options; ambiguous and matched known-risk
urns; willingness to accept and willingness to pay; sooner-versus-later
payments; gain, mixed, and loss lotteries; and advantageous and disadvantageous
allocation choices. Related screens often repeat a construct under a different
frame or payoff grid.

\begin{table}[t]
\centering
\caption{Multiple price list battery used as prediction targets}
\label{tab:task-map}
\small
\renewcommand{\arraystretch}{1.10}
\begin{tabularx}{\textwidth}{@{}>{\raggedright\arraybackslash}p{0.21\textwidth}
  >{\raggedright\arraybackslash}p{0.47\textwidth}X@{}}
\toprule
Domain & Raw task content & Economic variation \\
\midrule
Known risk & Certain amounts versus lotteries; matched known-risk urns & Probabilities and payoff spreads \\
Ambiguity & Ambiguous urns versus known-risk alternatives & Missing composition information \\
Compound lotteries & One-stage versus compound resolutions & Reduction of compound lotteries \\
Probability weighting & Common-ratio certain and lottery screens & Probability scale and common ratios \\
Time discounting & Sooner versus later payments & Delay and payment amount \\
Valuation & Willingness to accept and willingness to pay & Endowment and elicitation frame \\
Losses & Gain, mixed gain-loss, and loss lotteries & Sign and reference point \\
Distribution & Advantageous and disadvantageous allocations & Own and other's payoffs \\
\bottomrule
\end{tabularx}
\begin{minipage}{\textwidth}
\footnotesize
\textit{Notes:} The raw screens underlie 12 of the 21 econographics in
\citet{ChapmanEtAl2023}. Their published constructs transform switch intervals
into economic units and combine repeated elicitations. The outcomes here are
raw switch codes, not those measurement-error-corrected constructs.
\end{minipage}
\end{table}

The profile contains age, sex, race and ethnicity, education, income,
employment, household structure, region, religious attendance, six
International Cognitive Ability Resource items, three Cognitive Reflection
Test items, and metacognitive performance and confidence measures. These
variables are potentially useful, but they are not the main informational
object. The choices-only and profile-only designs separate their contribution
from the joint behavioural record. The processed replication file does not
contain the anxiety, depression, wellbeing, or personality scales sometimes
associated with richer survey panels.

The original split contains 800 training, 100 validation, and 100 test
participants. The confirmatory whole-domain result uses the test sample once
under a protocol fixed before those outcomes were loaded. Complete-sample
two-fold analyses use all 1,000 participants only for descriptive final
estimation and mechanism checks. They increase precision but do not replace
the one-time held-out estimate. All bootstrap intervals resample participants
and keep every task for a sampled participant together.

Appendix~\ref{app:identification-measurement} states formally how the raw-screen
prediction estimand differs from the authors' constructed-measure estimand and
which cross-domain restrictions distinguish a unified model from a collection
of target-specific regressions.

\section{Random Utility over Choice Embeddings}
\label{sec:embedded-utility}

\subsection{Representation and valuation}

The model has the three components in Figure~\ref{fig:unifying-architecture}.
The first constructs a common learned choice domain from the problem,
behavioural history, and feasible response. The second assigns one systematic
utility function on that domain. The third adds nonsystematic utility shocks
and maps total utility into stochastic choice. The first component supplies
expressive coordinates; only the second assigns stable economic value.

Let $H_{i,-t}$ collect decision maker $i$'s visible choices outside target task
$t$, let $Z_t$ describe the target problem, and let $\mathcal C_t$ index the
labelled peers used as context. The context table is
\[
 \mathcal D_t=\{(X_j,H_{j,-t},Z_t,Y_{jt}):j\in\mathcal C_t\}.
\]
For a whole domain mask, every occurrence of $H_{i,-t}$ is replaced by
$H_{i,-d(t)}$, so that no choice in task $t$'s domain $d(t)$ remains visible.
The vector $\omega\in\Omega$ contains the parameters of the public frozen
checkpoint. Its representation map produces
\begin{equation}
 h_{it}=\Phi_{\omega}(X_i,H_{i,-t},Z_t,\mathcal D_t)\in\mathbb R^p,
 \label{eq:foundation-state}
\end{equation}
where $p=2{,}048$ and $\omega$ remains fixed throughout the analysis.

Let $\Pi\in\mathbb R^{k\times p}$ be a fixed projection, let
$\alpha\in\mathbb R$, let $\beta\in\mathbb R^k$, and define
$\Lambda(v)=(1+\exp(-v))^{-1}$. One common valuation index maps the learned
state into the unit interval,
\begin{equation}
 g_{it}=\alpha+\beta^\top\Pi h_{it},
 \qquad \mu_{it}=\Lambda(g_{it}).
 \label{eq:embedded-utility}
\end{equation}
Let $\theta=(\alpha,\beta)$ collect the parameters of the valuation index. The
same $\theta$ applies to every decision maker, task, and domain. Individual
heterogeneity and the economic content of the problem enter through $h_{it}$;
the valuation parameters remain common. The scalar $\mu_{it}$ is the decision
maker's ideal normalized response plan for task $t$. It is an input to utility,
not utility itself.

The multiple price list interface restricts each observed response to one
switch. A feasible switch code $r\in\mathcal R_t$ therefore describes a
complete response plan for task $t$. Let $K_t$ be the number of binary rows,
let $c_t(r)$ be the number of option A choices implied by plan $r$, and define
$s_t(r)=c_t(r)/K_t$. The learned state and normalized response plan jointly
form the common choice object
\begin{equation}
 e_{it}(r)=\bigl(h_{it},s_t(r)\bigr)
 \in\mathcal M\equiv\mathbb R^p\times[0,1].
 \label{eq:common-choice-domain}
\end{equation}
The space $\mathcal M$ is identical across tasks and domains. Its first
coordinate retains the rich context of the problem and decision maker; its
second places every feasible response plan on a comparable scale. Let
$\kappa=1$ normalize the common utility scale. One systematic utility function values
every feasible object in this learned domain:
\begin{equation}
 V_{\theta}\!\left(e_{it}(r)\right)
 =-\kappa\left|s_t(r)-\mu_{it}\right|.
 \label{eq:plan-systematic-utility}
\end{equation}
This utility rule is identical across domains. The target problem changes the
learned state and the feasible response set, but it does not introduce a risk
coefficient, time coefficient, or distributional choice coefficient outside
the common index. Equation~\eqref{eq:plan-systematic-utility} is the economic
valuation of the observed response plans: utility is highest at the feasible
plan closest to $\mu_{it}$ and falls at the common normalized rate $\kappa=1$ with distance
from that ideal. Thus $V_{\theta}$, not $h_{it}$, $s_t(r)$, or $\mu_{it}$, is
systematic utility and ranks the observed menu.

The third component adds nonsystematic error on the same common valuation
scale. Let total utility be
$u_{it}(r)=V_{\theta}(e_{it}(r))+\varepsilon_{it}(r)$, where the disturbances
are independent standard type I extreme value draws across feasible response
plans. The resulting random utility model is
\begin{equation}
 \Pr(Y_{it}=r\mid X_i,H_{i,-t},Z_t,\mathcal D_t)
 =\frac{\exp\{V_{\theta}(e_{it}(r))\}}
 {\sum_{r'\in\mathcal R_t}\exp\{V_{\theta}(e_{it}(r'))\}}.
 \label{eq:plan-random-utility}
\end{equation}
Its modal response is
\begin{equation}
 \widehat Y_{it}^{\mathrm{hybrid}}
 =\arg\min_{r\in\mathcal R_t}
 \left|s_t(r)-\mu_{it}\right|,
 \label{eq:embedded-decoder}
\end{equation}
with ties resolved in favour of the smallest switch code. This is exactly the
nearest feasible decoder evaluated below. The point prediction tests the
systematic ranking; relaxing the normalization and calibrating $\kappa$ would
change probabilities but not the reported modal forecasts.

The same model also gives a direct account of cross-task dependence. Let
$Q_{it}=s_t(Y_{it})$ be the normalized observed response and let
$\mathcal H_i=(h_{i1},\ldots,h_{iT})$ collect the learned states for decision
maker $i$. Define $m_{it}=\mathbb E[Q_{it}\mid\mathcal H_i]$ and the residual
$\xi_{it}=Q_{it}-m_{it}$. For two tasks $t$ and $s$,
\begin{align}
 \operatorname{Cov}(Q_t,Q_s)
 &=\operatorname{Cov}(m_t,m_s)
 +\operatorname{Cov}(m_t,\xi_s)+\operatorname{Cov}(\xi_t,m_s)
 +\operatorname{Cov}(\xi_t,\xi_s).
 \label{eq:correlation-decomposition}
\end{align}
The two cross terms vanish by iterated expectations. If the remaining
cross-task shocks are also conditionally orthogonal, the residual covariance
vanishes. Correlation across domains is then generated by variation in one
systematic utility map evaluated at different learned task states. The
empirical test below substitutes task-specific cross-fitted systematic
forecasts for $m$ and reports every term rather than imposing orthogonality.

Equations \eqref{eq:foundation-state}--\eqref{eq:plan-random-utility} impose an
economic restriction: one learned choice domain, systematic utility function,
and shock distribution serve every domain. This parallels ordered random utility,
where restrictions on the map from types to utilities carry empirical content
\citep{ApesteguiaBallester2025}. Here the learned state is not itself utility;
equations \eqref{eq:embedded-utility} and
\eqref{eq:plan-systematic-utility} assign value. In the action-level extension,
familiar behavioural parameters would correspond to restrictions on slopes,
curvature, or asymmetries of the composite map from primitives through
$\Phi_\omega$ to $V$.

\subsection{Estimation and the cross domain restriction}

I estimate the common index using only the 800 training participants. The
input is TabFM's $p=2{,}048$ dimensional contextual state immediately before
its nonlinear decoder. The entries of $\Pi$ are drawn independently from
$N(0,1/k)$ using locked seed 20,260,806 and then held fixed; the projection
maps the state to $k=128$ coordinates. Let $C_{it}=K_tA_{it}$ be the number of
rows on which decision maker $i$ chooses option A. Define the continuity
corrected share $\widetilde A_{it}=(C_{it}+1/2)/(K_t+1)$ and target
\[
 \widetilde y_{it}=\log\!\left(
 \frac{\widetilde A_{it}}{1-\widetilde A_{it}}\right)
 =\log\!\left(\frac{C_{it}+1/2}{K_t-C_{it}+1/2}\right).
\]
For an estimation sample $\mathcal E_{\mathrm{fit}}$, the common index solves
\begin{equation}
 (\widehat\alpha,\widehat\beta)=
 \arg\min_{a\in\mathbb R,\,b\in\mathbb R^k}
 \sum_{(i,t)\in\mathcal E_{\mathrm{fit}}}
 \left(\widetilde y_{it}-a-b^\top\Pi h_{it}\right)^2
 +\lambda\lVert b\rVert_2^2.
 \label{eq:embedded-utility-estimator}
\end{equation}
Participant-grouped four-fold cross-validation selects $\lambda$ from
$\{0.01,0.1,1,10,100,1000\}$ by mean absolute error on the corrected option A
share. The intercept is not penalized. In each outer fold, the index is fitted
on 400 participants and predicts the other 400. The TabFM backbone remains
frozen and the target decision maker's response is absent from the
representation. Within each outer fold, one coefficient vector is common to
all 26 tasks; the two fold-specific estimates are used only to keep every
reported prediction out of sample by decision maker. Equation
\eqref{eq:embedded-utility-estimator} estimates the systematic index for point
prediction. The modal forecasts do not identify $\kappa$, so the analysis does
not claim calibrated choice probabilities.

The first test estimates one valuation index using all domains. The second
removes every outcome from one domain while fitting the index and then predicts
that domain in the opposite participant fold. The frozen representation still
uses the same 100 labelled target task examples supplied in the main design.
This is therefore a direct test of whether valuation transfers across domains,
holding the learned representation and its information set fixed.

The two layers do not have the same learning status. The foundation parameters
$\omega$ remain fixed while the context $\mathcal D_t$ changes the state and
prediction. By contrast, $(\alpha,\beta)$ are estimated from human choice data.
The results therefore establish the existence and portability of one utility
map once estimated; they do not show that the economic layer learns utility in
context.

A universal structural version would make that learning rule explicit. Let
$\mathcal Q_{it}$ collect the query's visible behavioural history and its labelled
context, but not $Y_{it}$. A fixed valuation learner $\mathcal A_\eta$ would
produce
\begin{equation}
 \theta_{it}=\mathcal A_\eta(\mathcal Q_{it}),
 \qquad
 V_{it}(r)=V\!\left(r,h_{it};\theta_{it}\right),
 \label{eq:in-context-valuation}
\end{equation}
where $\eta$ is pretrained and unchanged in the application. Context would
update the econometrician's inference about valuation, not the definition of
utility. The same learner $\mathcal A_\eta$, utility family $V$, and shock law
would serve every domain. The current estimator is not this in-context
valuation learner; equation~\eqref{eq:in-context-valuation} states the gap
between portable estimated utility and the foundation model's universal
adaptation.

The current welfare object covers observed response plans. Appendix
\ref{app:primitive-action} gives the action-level extension needed for new
payoff menus and the formal Lean verification of the architecture.

\section{Empirical Design}

\subsection{A frozen foundation model with labelled context}

TabFM is a pretrained regression and classification model for mixed-type tables
\citep{KongDas2026,GoogleResearchTabFM2026}. It alternates attention across
columns and rows, compresses each row into a learned representation, and uses a
24-block in-context transformer to map labelled context rows and unlabelled query
rows into predictions. The regression checkpoint has a scalar decoder.
Appendix~\ref{app:tabfm-economists} explains the architecture and public
disclosure in terms familiar to economists.

All primary model parameters remain frozen. This is parameter-zero-shot,
in-context supervised prediction. In-context learning (ICL) means that labelled
examples enter the model's input at prediction time: no Econographics gradient
update or model-specific hyperparameter search occurs, while the model observes
outcomes for other participants. The stricter zero-label exercise removes every
label from the target task or domain.

For each target $t$, I build a regression episode. Its context is 100 other
participants with features $(X_j,Y_{j,-t},Z_t)$ and label $Y_{jt}$. The query
has the same feature schema but not the target response. Two deterministic
participant folds are balanced in size, and every query uses context rows from
the opposite fold. The focal participant is therefore never among the labelled
examples used to predict them. Context identities are fixed functions of task,
fold, and a stored seed.

Routing provides 100 labels for the exact target. Within an episode, task
descriptors that are constant across context rows identify the episode and its
feasible response map but may be removed by preprocessing. Cross-participant
variation must therefore come from the profile, visible choices, and their
relationship with the target labels. The setup asks whether one pretrained
inference engine can learn each local mapping from a coherent behavioural
history. It is more demanding than estimating a population median, but less
demanding than predicting a wholly unseen task without local labels.

\subsection{Two masking designs}

The task mask hides one of 26 responses and leaves the other 25 visible. Every
participant is queried for every target, producing 26,000 cross-fitted
predictions. Because a related within-domain screen often remains visible,
this is primarily a reconstruction design. It measures how much information is
contained in the complete behavioural vector and supports high-powered
falsification tests.

The whole-domain mask removes every task in one of eight domains. An outcome-
blind Secure Hash Algorithm 256-bit (SHA-256) permutation assigns exactly one
domain to each participant in the original split. The 800 training
participants contribute 100 assignments per domain; the validation and test
groups contribute 12 or 13. A query for domain $d$ contains the profile and
choices in the other seven domains, while every outcome in $d$ is missing.

Validation was used to choose task routing and domain-specific shrinkage toward
the training-sample median. Before the test was evaluated, a version-controlled
protocol fixed the participants, contexts, estimator, shrinkage coefficients,
seeds, metrics, and baselines. The 100 test participants were then evaluated
once. Later full-sample cross-fits report the raw frozen output as their primary
descriptive estimate; using the validation-selected shrinkage after including
validation participants would mix selection and estimation.

\subsection{Information interventions and matched learners}

Table~\ref{tab:falsifications} summarizes the main tournament. Participant
folds, targets, context identities, and feasible response maps never change.
Only the query information, the validity of its decision-maker linkage, the
validity of context labels, or the prediction rule changes.

\begin{table}[t]
\centering
\caption{Prediction tournament}
\label{tab:falsifications}
\small
\setstretch{1.00}
\renewcommand{\arraystretch}{1.13}
\begin{tabularx}{\textwidth}{@{}>{\raggedright\arraybackslash}p{0.21\textwidth}
  >{\raggedright\arraybackslash}p{0.45\textwidth}
  >{\raggedright\arraybackslash}X@{}}
\toprule
Specification & Information supplied & Comparison isolated \\
\midrule
Frozen foundation model (TabFM) & Profile, 25 visible choices, target features, and 100 context rows & Primary frozen specification \\
Choices only & All visible choices; profile removed & Incremental value of profile data \\
Profile only & Demographics, cognition, and metacognition; choices removed & Profile without behavioural history \\
Five-choice history & Profile and five outcome-blind visible tasks & Value of a dense history \\
Shuffled decision maker & Independently participant-shuffled visible task values & Same-decision-maker coherence, holding task marginals fixed \\
Shuffled labels & True features with target labels shuffled within task and fold & Valid in-context feature-label mapping \\
Ridge, Extra Trees, and CatBoost & Full features and the same 100 context rows & Matched linear and nonlinear fits \\
Context median & Same 100 target labels; no query covariates & Matched population forecast \\
\bottomrule
\end{tabularx}
\begin{minipage}{\textwidth}
\footnotesize
\textit{Notes:} Multiple price list responses are always hidden for the target.
The decision-maker-shuffle intervention preserves each task's marginal distribution
but makes the visible vector incoherent. Classical learners are refitted inside
each episode; TabFM receives the same labels in its input and keeps its weights
fixed.
\end{minipage}
\end{table}

The decision-maker-shuffle and shuffled-label interventions address different
shortcuts. The first asks whether a bag of plausible marginal responses is
enough. The second asks whether TabFM learns the local mapping from the new
table or simply transforms a query into a fixed score. Choices only tests
whether the rich profile drives the result. Five choices test how performance
changes when the behavioural history becomes sparse.

The primary nonlinear benchmark is CatBoost. Modern tabular benchmarks find
that boosted trees remain strong, while cross-model ensembles lead overall and
foundation models are especially competitive on smaller tables
\citep{EricksonEtAl2025TabArena,HollmannEtAl2025}. CatBoost is a single,
reproducible representative of that frontier. It handles the mixed categorical
table directly and uses ordered boosting to limit target leakage
\citep{ProkhorenkovaEtAl2018}. For every target and fold, I select depth,
learning rate, regularization, and iteration count by three-fold validation
within the 100 context rows. Query outcomes never enter model selection.

Extra Trees remains a secondary computational diagnostic
\citep{GeurtsErnstWehenkel2006}. It captures nonlinear interactions and is
cheap to refit for every episode. Together the two learners separate a result
about the information in the table from a result about the frozen inference
rule. Ridge supplies the corresponding linear benchmark.

The primary point loss averages absolute row error within participant. I also
report normalized MAE, exact accuracy, and accuracy within one and two rows.
Paired confidence intervals resample participants 10,000 times and treat tasks
as fixed. The participant distribution in equation~\eqref{eq:participant-fitness}
is shown directly, rather than reducing every comparison to a mean.

\subsection{Locked internal-representation test}

A secondary train-only design traces all 26 tasks through the frozen
transformer. After a ten-task pilot, I locked the domain labels, 200
participants, two folds, one projection, and 99,999 label permutations. The
primary statistic is the input-to-output change in the within-domain minus
between-domain gap in linear centred-kernel alignment (CKA). Cross-fitted
probes measure when predictions and responses become recoverable.
Appendix~\ref{app:layerwise-lens} gives the full protocol and audit.

\section{Results}
\suppressfloats[t]

\subsection{The held-out domain test}

The one-time held-out result covers 100 participants and 324 tasks in their
outcome-blind assigned domains. Table~\ref{tab:heldout-point} reports levels and
paired contrasts for the locked validation-shrunken forecast.

\begin{table}[H]
\centering
\caption{Held-out prediction of a randomly hidden domain}
\label{tab:heldout-point}
\small
\setlength{\tabcolsep}{4.0pt}
\begin{tabularx}{\textwidth}{@{}>{\raggedright\arraybackslash}p{0.38\textwidth}*{3}{>{\centering\arraybackslash}X}@{}}
\toprule
& Row MAE & Normalized MAE & Within 2 (\%) \\
\midrule
\multicolumn{4}{l}{\textit{Panel A. Forecast levels}} \\
Frozen TabFM, locked forecast & 3.538 & 0.227 & 50.0 \\
Training-sample population median & 4.285 & 0.273 & 40.5 \\
TabFM, decision-maker shuffle & 4.548 & 0.288 & 32.0 \\
\addlinespace
\multicolumn{4}{l}{\textit{Panel B. TabFM improvement and 95 percent interval}} \\
Versus population median & 0.748 & 0.046 & 9.5 \\[-2pt]
& {\scriptsize $[0.303,1.208]$} & {\scriptsize $[0.019,0.075]$} & {\scriptsize $[2.8,16.5]$} \\[1pt]
Versus shuffled decision maker & 1.010 & 0.062 & 18.0 \\[-2pt]
& {\scriptsize $[0.395,1.645]$} & {\scriptsize $[0.024,0.101]$} & {\scriptsize $[8.8,27.0]$} \\
\bottomrule
\end{tabularx}
\begin{minipage}{0.95\textwidth}
\footnotesize
\textit{Notes:} There are 100 held-out participants and 324 hidden task
outcomes. MAE denotes mean absolute error. Normalized MAE divides row error by
the feasible row span before averaging within participant. Positive values in
Panel B favour TabFM. Intervals use 20,000 participant bootstrap draws. The
decision-maker-shuffle condition preserves every visible task marginal but reassigns
task values independently across participants.
\end{minipage}
\end{table}

TabFM lowers row MAE by 17.4 percent relative to the population median and
raises within-two accuracy by 9.5 percentage points. Reassigning the visible
history is more damaging: the TabFM advantage rises to 1.010 rows and 18.0
points. The hidden domain contributes no query outcome, all comparison
quantities come from training participants, and assignment is outcome blind.
The result therefore identifies portable information about the same decision maker, not a bag of
plausible marginal responses.

The raw unshrunk forecast has lower ex post test MAE, 3.305, but was not the
locked specification. It remains a sensitivity. Complete-sample cross-fits use
the raw model for descriptive estimation and do not replace this test.

\subsection{One valuation rule across domains}

Table~\ref{tab:embedding-utility} reports the main hybrid result. The common
valuation index and the foundation representation form the systematic part of
the random-utility specification in equations \eqref{eq:embedded-utility}--
\eqref{eq:plan-random-utility}. The reported modal forecasts test its ranking
and do not estimate the common shock scale. All predictions are out of sample
by decision maker, and the foundation checkpoint remains frozen.

\begin{table}[H]
\centering
\caption{Random utility in the learned choice space}
\label{tab:embedding-utility}
\small
\setlength{\tabcolsep}{4.0pt}
\begin{tabular}{lrrrr}
\toprule
Prediction rule & Row MAE & Normalized MAE & Exact (\%) & Within 2 (\%) \\
\midrule
TabFM nonlinear decoder & 2.146 & 0.125 & 23.14 & 71.04 \\
Common utility index & 2.429 & 0.141 & 15.25 & 65.83 \\
Leave one domain out utility index & 2.550 & 0.149 & 13.97 & 63.16 \\
Task population median & 4.067 & 0.234 & 12.07 & 40.42 \\
\bottomrule
\end{tabular}
\begin{minipage}{0.95\textwidth}
\footnotesize
\textit{Notes:} The sample contains 20,800 train only decision maker by task
predictions. MAE denotes mean absolute error. Within each participant fold, the
common utility index uses one intercept, one coefficient vector, and one
response-plan utility across all 26 tasks. The two fold-specific estimates keep
predictions out of sample. The leave one domain out row also excludes the target domain's outcomes
when fitting the valuation index. The frozen representation retains the 100
target task context labels used in the main design. The population median is
estimated in the opposite participant fold. Validation and test outcomes are
not loaded.
\end{minipage}
\end{table}

The common valuation rule recovers most of the foundation model's predictive
gain. Its within two accuracy is 65.83 percent, compared with 40.42 percent for
the population median and 71.04 percent for TabFM's nonlinear decoder. On
normalized error, the hybrid retains 85 percent of TabFM's improvement over
the population forecast. Its participant clustered normalized error difference
relative to the population median is $-0.0927$, with 95 percent interval
$[-0.0984,-0.0871]$.

The stronger valuation transfer restriction remains predictive. After every
target domain outcome is excluded from estimation, the same index form reaches
63.16 percent within two rows and normalized error of 0.149. It preserves 78
percent of the full decoder's normalized error improvement over the population
median. The same valuation direction learned from the other seven domains
therefore predicts the eighth. The result locates most of the model's unity in
a common valuation rule over the learned representation rather than in 26
separate nonlinear decoding rules.

\subsection{One utility map reproduces joint behaviour}

Passing the cross-fitted forecasts through the exact public construct
transformations and measurement correction in \citet{ChapmanEtAl2023} yields a
sharper result. Across 66 correlations among 12 constructs, the common-utility
matrix has Pearson alignment 0.959 and rank alignment 0.939 with the observed
matrix; a 9,999-draw quadratic-assignment test gives $p=0.0001$. Fitting the
utility index without target-domain outcomes gives rank alignment 0.935,
compared with 0.870 for TabFM's nonlinear decoder. TabFM is therefore the
predictive unified model, and random utility is an economic compression that
preserves both its accuracy and the ordering of the Chapman dependence
pattern. It overstates mean absolute off-diagonal correlation, so the claim
concerns the geometry of dependence rather than its complete level. Appendix
Table~\ref{tab:chapman-utility-matrix} reports the covariance decomposition and
participant bootstrap.

\subsection{Complete-sample reconstruction and falsification}

The task-mask analysis produces 26,000 out-of-fold predictions: 26 targets for
each of 1,000 participants. Table~\ref{tab:all-ablations} reports the complete
tournament. Frozen TabFM is exactly correct on 23.05 percent of choices, within
one row on 52.86 percent, and within two rows on 69.91 percent. Its row MAE is
2.185. The matched context median reaches 38.58 percent within two rows and has
MAE 4.101. A uniform distribution over feasible rows has expected within-two
accuracy of 27.03 percent and MAE of 5.775.

\begin{table}[t]
\centering
\small
\setlength{\tabcolsep}{3.4pt}
\caption{Complete-sample prediction tournament}
\label{tab:all-ablations}
\IfFileExists{tables/full_sample_ablations.tex}{\begin{tabular}{lrrrr}
\toprule
Predictor & Exact (\%) & Within 1 (\%) & Within 2 (\%) & Row MAE \\
\midrule
Frozen \TabFM{}                   & 23.05 & 52.86 & 69.91 & 2.185 \\
Extra Trees                       & 20.12 & 49.44 & 67.73 & 2.282 \\
CatBoost                          & 16.27 & 44.35 & 64.40 & 2.426 \\
Ridge                             & 11.37 & 32.60 & 52.00 & 2.927 \\
Five task history                 & 10.04 & 28.06 & 44.59 & 3.494 \\
Matched 100-participant context median & 12.33 & 27.42 & 38.58 & 4.101 \\
Shuffled decision maker              &  5.91 & 18.62 & 32.12 & 4.248 \\
Shuffled context labels           &  5.96 & 18.29 & 31.98 & 4.246 \\
Profile only                      &  6.03 & 18.10 & 31.58 & 4.226 \\
Choices only                      & 22.90 & 51.97 & 69.15 & 2.216 \\
Uniform feasible-row expectation  &  6.06 & 17.12 & 27.03 & 5.775 \\
\bottomrule
\end{tabular}
}{%
\begin{tabular}{lrrrr}
\toprule
Predictor & Exact & Within 1 & Within 2 & Row MAE \\
\midrule
\multicolumn{5}{c}{Generated table unavailable} \\
\bottomrule
\end{tabular}}
\begin{minipage}{0.95\textwidth}
\footnotesize
\textit{Notes:} There are 1,000 participants, 26 targets, and two deterministic
participant folds. Every query uses the same 100 context decision makers from the
opposite fold. TabFM weights remain frozen. Extra Trees and Ridge are refitted
on the displayed context. CatBoost is tuned by three-fold validation within
those context rows and then refitted on all 100. Query outcomes never enter
tuning. Complete-sample estimates are descriptive final estimation and do not
replace the held-out test.
\end{minipage}
\end{table}

The information interventions identify what drives this performance. Choices
alone retain 69.15 percent within-two accuracy, only 0.75 percentage points
below the complete profile. Profile alone falls to 31.58 percent, below the
matched population median. Five outcome-blind choices reach 44.59 percent, so
even a sparse history is useful, but it remains far below the 25-choice vector.
The main information source is the behavioural record rather than demographics,
cognition, or metacognition.

Both destructive interventions collapse. Accuracy under the decision-maker shuffle is 32.12
percent within two rows and shuffled-label accuracy is 31.98 percent. The first
destroys the coherent decision-maker vector while preserving each choice marginal. The
second preserves context features and label marginals while destroying their
local mapping. TabFM therefore uses a coherent behavioural history and valid
in-context labels. It does not emit a fixed behavioural score from the query
alone.

Two locally trained nonlinear learners sharpen the comparison. Extra Trees
reaches 67.73 percent within two rows and MAE 2.282. Frozen TabFM is higher by
2.18 percentage points, with participant-clustered 95 percent interval
$[1.73,2.63]$, and lowers MAE by 0.097 rows, interval $[0.081,0.113]$. Tuned
CatBoost reaches 64.40 percent within two rows and MAE 2.426. The corresponding
TabFM advantages are 5.50 percentage points, interval $[4.96,6.07]$, and 0.241
rows, interval $[0.221,0.261]$. Ridge is substantially weaker. Most predictive
content is therefore accessible to a flexible classical learner, but the
frozen foundation model retains a precisely estimated advantage under the same
features and labels.

\subsection{Who benefits from the unified predictor?}

\begin{figure}[t]
\centering
\includegraphics[width=\textwidth]{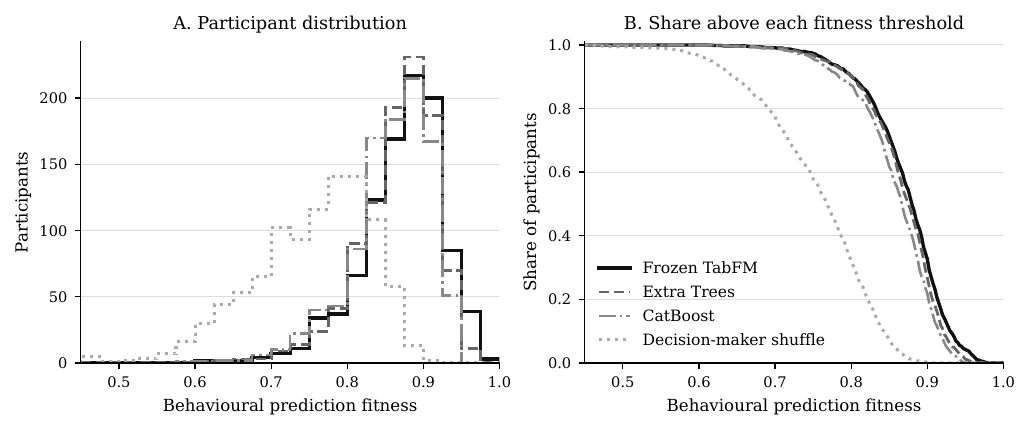}
\caption{Participant-level behavioural prediction fitness}
\label{fig:participant-fitness}
\begin{minipage}{0.96\textwidth}
\footnotesize
\textit{Notes:} Fitness is one minus participant-mean absolute error on the
normalized share-of-A-choices scale. Each distribution contains 1,000
participants and averages over 26 cross-fitted task predictions. Extra Trees
and CatBoost use the same features and 100 context rows as TabFM. The
decision-maker shuffle independently reassigns every visible task across participants,
preserving each task marginal but breaking the coherent decision-maker vector.
\end{minipage}
\end{figure}

Figure \ref{fig:participant-fitness} shows whether gains are broad or
concentrated. Mean fitness is 0.871 for frozen TabFM, 0.866 for Extra Trees,
0.858 for CatBoost, and 0.754 after shuffling decision maker identity. TabFM
improves on these benchmarks for 652, 788, and 988 of 1,000 participants,
respectively. Its survival curve lies to the right of both local learners over
most of the relevant range and far to the right of the falsification.

The advantage over a strong nonlinear learner reaches almost two thirds of
participants. The larger decision maker coherence effect is nearly universal.

\subsection{The harder complete-sample domain mask}

Hiding each of eight domains for every participant produces 26,000 task
predictions. Frozen TabFM lowers row MAE from 4.056 for the matched context
median to 3.455 and raises within two accuracy from 40.26 to 45.81 percent. It
improves MAE in seven domains and within two accuracy in six. Concentrated time
discounting responses favour the population forecast, as do distributional
choices under the within two loss. The aggregate advantage is a stable error
reduction when every local within domain bridge is removed. Appendix
Tables~\ref{tab:zero-label} and~\ref{tab:heldout-joint} report the stricter
zero label and joint distribution exercises.

\section{\hspace{0.30em}The Geometry of Joint Behaviour}
\label{sec:geometry-main}

Prediction establishes that choices outside a domain contain information about
choices inside it. A unified model should also organize decision makers in a
common space. If two tasks are behaviourally near, the same decision makers
should occupy similar relative positions when either task is predicted. Let
$E_t$ collect the projected contextual states of the same decision makers
queried on task $t$, with rows aligned by decision maker. Let
$\widetilde E_t$ subtract the task-by-fold column means from $E_t$. I measure
task nearness by linear centred-kernel alignment (CKA),
\begin{equation}
 \operatorname{CKA}(t,s)=
 \frac{\lVert \widetilde E_t' \widetilde E_s\rVert_F^2}
 {\sqrt{\lVert \widetilde E_t'\widetilde E_t\rVert_F^2
 \lVert \widetilde E_s'\widetilde E_s\rVert_F^2}}.
 \label{eq:cka}
\end{equation}
This statistic is invariant to orthogonal rotation and isotropic rescaling. It
therefore compares participant geometry rather than individual coordinates.

\begin{figure}[H]
\centering
\includegraphics[width=0.96\textwidth]{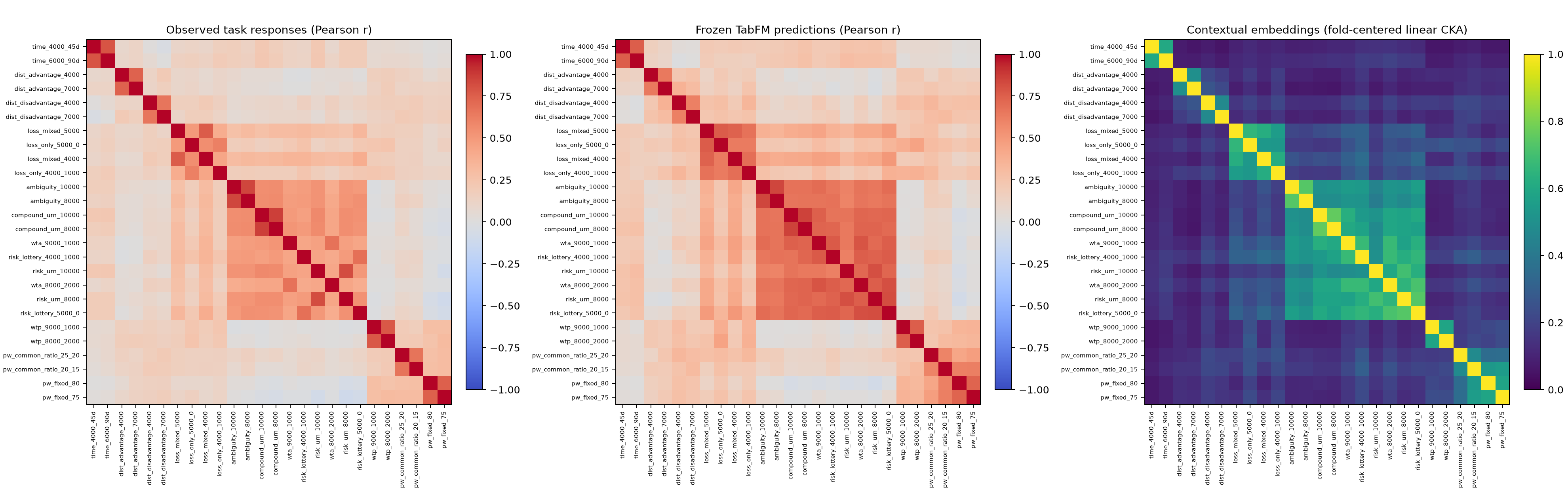}
\caption{Observed, Predicted, and Represented Task Nearness}
\label{fig:task-nearness}
\begin{minipage}{0.94\textwidth}
\footnotesize
\textit{Notes:} All panels use the same hierarchical task order. The first two
panels are Pearson correlations across 1,000 decision makers. The third is
fold-centred linear CKA over projected contextual states. CKA ranges from zero
to one; correlations range from minus one to one.
\end{minipage}
\end{figure}

Figure~\ref{fig:task-nearness} compares observed behaviour, frozen predictions,
and internal representation. Mean contextual CKA is 0.487 for tasks in the
same economic group and 0.206 across groups. The difference exceeds all 5,000
task-label permutations (plus-one one-sided $p=0.0002$). The upper triangle of the CKA
matrix has rank correlation 0.769 with absolute observed response correlation
and 0.828 with absolute predicted correlation. The model therefore reconstructs
not only missing coordinates but also much of the dependence pattern among
tasks.

The geometry requires coherent information. Spearman correlation between the
upper triangles of the state-CKA and absolute observed-response correlation
matrices is 0.927 at the decoder state and 0.922 when only choices are visible.
It falls to 0.155 after shuffling decision makers, minus 0.118 after shuffling
context labels, and 0.088 when only the profile is visible. The common space is
therefore organized by the same decision maker's behavioural record and by a
valid mapping from context features to labels.

\subsection{Where the common geometry emerges}

A locked train-only extraction follows 200 decision makers and all 26 tasks
through the 24 in-context-learning blocks. The design, domain partition,
projection, participant folds, and 99,999-draw randomization test were fixed
after a ten-task pilot. No validation or test record enters this exercise.

\begin{figure}[H]
\centering
\includegraphics[width=0.74\textwidth]{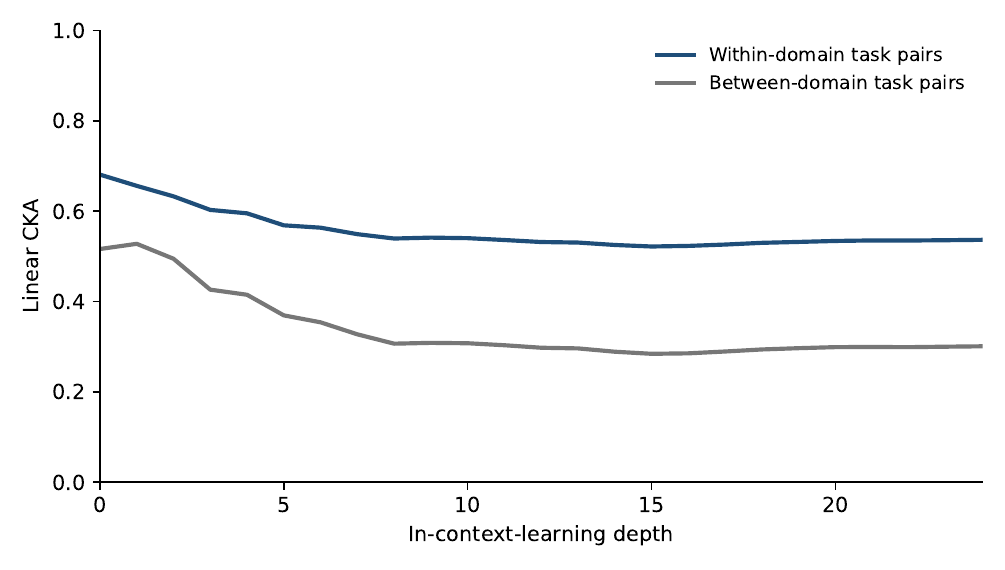}
\caption{Economic Domain Geometry Across In-Context-Learning Depth}
\label{fig:layerwise-domain-cka}
\begin{minipage}{0.90\textwidth}
\footnotesize
\textit{Notes:} Lines report mean CKA across decision makers for task pairs in
the same catalogue domain and in different domains. Fold means are removed by
target before CKA. The increase in the within-domain minus between-domain
contrast is the locked statistic.
\end{minipage}
\end{figure}

Figure~\ref{fig:layerwise-domain-cka} shows the internal transformation. At the
input, mean CKA is 0.681 within domains and 0.516 across domains, a contrast of
0.165. At the output, the values are 0.537 and 0.301, a contrast of 0.236. None
of 99,999 fixed-size label permutations produces an increase as large. Yet the
24-block update signatures have cosine similarity above 0.9997 across domains.
The model does not reserve risk blocks and time blocks. One computational
sequence carries every domain while increasingly separating unrelated tasks
and preserving more common decision maker geometry within related domains.

The final prediction becomes more linearly recoverable with depth, while the
observed response is already recoverable at the input and changes little. The
transformer computes its answer from behavioural information already present in
the common state. Appendix~\ref{app:layerwise-lens} reports the probe table,
forward-pass audit, and additional geometry checks.

\section{\hspace{0.30em}What Common Utility Changes}

The foundation model supplies a common language for unlike alternatives; the
estimated index values it. One common index retains 85 percent of the full
model's normalized-error improvement, 78 percent when the target domain does
not fit the index, and the ordering of the Chapman correlation matrix. TabFM is
the predictive unified model for the battery; random utility gives that
capability an economic form.

The matched synthetic adaptation clarifies what economic theory contributes.
Marginal theories help after training-sample level alignment, but preserving
their synthetic cross-domain links does not improve on independently permuted
domain blocks at equal compute.

The two layers learn differently. TabFM adapts from labelled context with
frozen weights; the economic index is estimated and then transported. Equation
\eqref{eq:in-context-valuation} defines the synthesis in which a fixed economic
learner infers utility from context. Welfare on new menus additionally requires
the action representation in equation~\eqref{eq:action-embedding-rum} to remain
stable.

\section{\hspace{0.30em}Related Literature}

Choice-based foundations restrict stochastic utility \citep{GulPesendorfer2006}.
Ordered random utility locates empirical content in the type-to-utility map
\citep{ApesteguiaBallester2025}. This paper adds cross-domain portability to
joint preference measurement \citep{ChapmanEtAl2023}. Appendix~\ref{app:literature}
gives the broader map.

\section{Conclusion}
Random utility supplies a common principle on an abstract domain. A foundation
model makes that domain expressive; one systematic utility function values it;
and one shock law generates stochastic choice. The estimated map transfers
across domains and organizes their dependence. Learned representation and
economic valuation can therefore form one unified model without conflating
prediction, utility, and noise.

\begingroup
\sloppy
\bibliographystyle{aer}
\bibliography{references}
\endgroup

\clearpage
\appendix
\section{Data Construction}
\label{app:formal-data}

\subsection{Profiles, supports, and missingness}

Every target has an explicit feasible switch support. Predictions are made on
the share-of-A-choices scale and projected to the nearest feasible code. This
avoids treating a never-switch response as an ordinary interior row. Raw row
MAE remains useful for interpretation, while normalized MAE permits comparison
across tasks with different grid lengths.

Profile features are built only from fields present in the processed public
replication file. Continuous variables are passed numerically; categorical
variables are encoded by the TabFM wrapper or the matched classical pipeline.
Missing profile values are imputed from context rows. Target outcomes are never
imputed and are absent from every query. Audit files record participant and
task counts, context identities, feasible maps, missing prediction counts, and
whether test profiles, actions, or labels were loaded.

The CatBoost benchmark receives the same profile, visible choices, target
description, 100 labels, and query rows as the frozen model. Its categorical
columns remain categorical. Context-constant columns are removed. A fixed grid
crosses depths 2, 3, 4, and 6 with two learning-rate, regularization, and
iteration schedules. Three-fold context-only validation selects the
configuration separately for each target and participant fold. The final model
is refitted on all 100 context rows. The audit stores the package version,
candidate scores, selected configuration, context identities, and prediction
array hash.

\subsection{Three claims and their identifying comparisons}

This subsection fixes the estimand used to distinguish the paper's three claims.
Let $\mathcal E$ denote a fixed evaluation design: its participants, target
masks, context rows, and feasible-response maps. Let
$\mathcal I_{\mathcal E}$ be its participants, and let
$\mathcal T_i^{\mathcal E}$ be the masked tasks evaluated for participant $i$.
A predictor $f$ maps the information supplied by the specified design into a
predicted option-A share
$\widehat A_{it}(f)\in[0,1]$. Its participant-weighted normalized mean absolute
error is
\begin{equation}
 \mathcal L_{\mathcal E}(f)=
 \frac{1}{|\mathcal I_{\mathcal E}|}
 \sum_{i\in\mathcal I_{\mathcal E}}
 \frac{1}{|\mathcal T_i^{\mathcal E}|}
 \sum_{t\in\mathcal T_i^{\mathcal E}}
 \left|A_{it}-\widehat A_{it}(f)\right|.
 \label{eq:evaluation-loss}
\end{equation}
Here $A_{it}$ is the observed option-A share defined in
Section~\ref{sec:prediction-data}, and every comparison holds $\mathcal E$, the target masks, context
identities, and feasible-response maps fixed. Let $f^{\mathrm{full}}$ denote
the frozen predictor with the complete visible history,
$f^{\mathrm{med}}$ the opposite-fold context median,
$f^{\mathrm{profile}}$ the profile-only predictor,
$f^{\mathrm{shuffle}}$ the predictor after independently shuffling each visible
choice coordinate across participants, and $f^{\mathrm{labels}}$ the predictor after
shuffling context labels within task and fold. Finally,
$f^{\mathrm{hybrid}}$ is the common valuation index and response plan random
utility rule estimated in Section~\ref{sec:embedded-utility}. These definitions make each row of
Table~\ref{tab:claim-levels} a
specific loss comparison rather than a verbal hierarchy.

\begin{table}[H]
\centering
\caption{Three claims, samples, and identifying comparisons}
\label{tab:claim-levels}
\small
\setstretch{1.00}
\renewcommand{\arraystretch}{1.14}
\begin{tabularx}{\textwidth}{@{}>{\raggedright\arraybackslash}p{0.18\textwidth}
  >{\raggedright\arraybackslash}p{0.31\textwidth}
  >{\raggedright\arraybackslash}p{0.27\textwidth}
  >{\raggedright\arraybackslash}X@{}}
\toprule
Claim & Formal comparison & Evaluation sample & Status \\
\midrule
Cross-domain signal &
$\mathcal L_{\mathcal E}(f^{\mathrm{full}})$ is below both
$\mathcal L_{\mathcal E}(f^{\mathrm{med}})$ and
$\mathcal L_{\mathcal E}(f^{\mathrm{profile}})$. &
Complete-sample cross-fit; the median contrast is also held out &
Supported \\
Frozen unified predictor &
$\mathcal L_{\mathcal E}(f^{\mathrm{full}})$ is below the losses from
decision-maker and label shuffling, $\mathcal L_{\mathcal E}(f^{\mathrm{shuffle}})$ and
$\mathcal L_{\mathcal E}(f^{\mathrm{labels}})$. &
Complete-sample cross-fit with the same checkpoint and contexts &
Supported \\
Common plan utility &
$\mathcal L_{\mathcal E}(f^{\mathrm{hybrid}})$ is below
$\mathcal L_{\mathcal E}(f^{\mathrm{med}})$, including when target-domain
outcomes are excluded from fitting the valuation index. &
800-participant training sample; participant cross-fit &
Supported for valuation transfer \\
\bottomrule
\end{tabularx}
\begin{minipage}{\textwidth}
\footnotesize
\textit{Notes:} ``Valuation transfer'' means that target-domain outcomes are
excluded when estimating $f^{\mathrm{hybrid}}$; the frozen representation still
uses the 100 target-task labels supplied to its in-context input. The first two
rows use participant-clustered inference.
\end{minipage}
\end{table}

\subsection{Synthetic-linkage contrast}

The intervention measures the predictive contribution of joining marginal
economic theories within that protocol. For synthetic replication $b$, let
$\mathcal S_b^{\mathrm{joint}}$ contain all domain blocks generated from the
latent parameters of the same simulated decision maker. Let $\pi_b$ independently permute
complete domain blocks across synthetic identities within each registered
generator stratum, and define
$\mathcal S_b^{\mathrm{shuf}}=\pi_b(\mathcal S_b^{\mathrm{joint}})$. This
operation preserves every domain-specific block and its empirical marginal
distribution while removing cross-domain identity. Starting from the same
pretrained checkpoint and paired schedule, define
\[
 f_b^{\mathrm{joint}}=
 \operatorname{Fit}(\omega^{\mathrm{pre}},\mathcal S_b^{\mathrm{joint}};\xi_b),
 \qquad
 f_b^{\mathrm{shuf}}=
 \operatorname{Fit}(\omega^{\mathrm{pre}},\mathcal S_b^{\mathrm{shuf}};\xi_b).
\]
Let $\mathbb P^{\mathrm{link}}$ be the registered law for the synthetic
generator, block permutation, and paired optimization schedule. The
synthetic-linkage contrast is
\begin{equation}
 \Delta_{\mathrm{join}}=
 \mathbb E_{b\sim\mathbb P^{\mathrm{link}}}\!\left[
 \mathcal L_{\mathcal E}(f_b^{\mathrm{shuf}})
 -\mathcal L_{\mathcal E}(f_b^{\mathrm{joint}})\right].
 \label{eq:joint-prior-value}
\end{equation}
A positive $\Delta_{\mathrm{join}}$ means that synthetic cross-domain
dependence improves expected prediction under the registered generator,
adaptation rule, and evaluation sample. It does not by itself identify a
structural preference parameter.

The equal-compute implementation finds no stable positive linkage effect on
human validation. The joint adapter lowers unanchored normalized error by
0.00371 relative to the independent-marginal adapter, with 95 percent interval
$[-0.00193,0.00937]$. After applying the identical training-sample marginal
anchor to both arms, the difference reverses to $-0.00368$, with interval
$[-0.00921,0.00185]$. Appendix Table~\ref{tab:structural-validation} reports the
levels. The current synthetic linkage therefore adds no detectable transfer
beyond the matched marginal economic theories.

\section{Identification and Measurement}
\label{app:identification-measurement}

The interpretation of the results turns on two distinctions. The first is
between running common software on many tasks and imposing a restriction that
links those tasks. The second is between forecasting the 26 raw multiple price
list screens and reconstructing the economic measures in
\citet{ChapmanEtAl2023}. This section states both distinctions formally and
collects the corresponding evidence.

\subsection{From a common algorithm to a common model}

Consider first an unrestricted collection of local prediction rules,
\begin{equation}
 \widehat A_{it}=f_t(X_i,H_{i,-t};\mathcal D_t),
 \qquad t=1,\ldots,T,
 \label{eq:local-stack}
\end{equation}
where $f_t$ may differ arbitrarily across targets. Using the same software to
estimate every $f_t$ does not restrict this collection. The 100 labels for the
exact target in the main TabFM episode can support a different local mapping
for every $t$, even though the checkpoint is fixed. Frozen weights are
therefore evidence of parameter portability, not by themselves evidence of a
common behavioural restriction.

The response-plan model in Section~\ref{sec:embedded-utility} imposes the
restriction missing from equation~\eqref{eq:local-stack}. Let $D_t$ map a
normalized option-A share to the option-A share of the nearest feasible
response plan for task $t$. Its forecast is
\begin{equation}
 \widehat A_{it}^{\mathrm{CU}}
 =D_t\!\left[
 \Lambda\!\left(\alpha+\beta^\top\Pi
 \Phi_{\omega}(X_i,H_{i,-t},Z_t,\mathcal D_t)\right)
 \right].
 \label{eq:common-model-audit}
\end{equation}
The checkpoint $\omega$, projection $\Pi$, coefficients $(\alpha,\beta)$,
plan-utility function, and disturbance law are common. Only the information
state and feasible response set change with the problem. For target domain
$d$, let $(\widehat\alpha_{-d},\widehat\beta_{-d})$ be estimated without any
outcome from $d$. Prediction with these coefficients is a direct cross-domain
restriction: the target domain cannot fit its own valuation map.

Three loss contrasts locate the empirical content of the model. Using the
evaluation loss $\mathcal L_{\mathcal E}$ defined in
equation~\eqref{eq:evaluation-loss}, define
\begin{align}
 \Delta_{\mathrm{history}}
 &=\mathcal L_{\mathcal E}(f^{\mathrm{shuffle}})
   -\mathcal L_{\mathcal E}(f^{\mathrm{full}}),
 &
 \Delta_{\mathrm{domain}}
 &=\mathcal L_{\mathcal E}(f^{\mathrm{med}})
   -\mathcal L_{\mathcal E}(f^{\mathrm{full}}),
 \label{eq:unity-contrasts}\\
 \Delta_{\mathrm{value}}
 &=\mathcal L_{\mathcal E}(f^{\mathrm{med}})
   -\mathcal L_{\mathcal E}(f^{\mathrm{CU}}_{-d}).
 \nonumber
\end{align}
A positive $\Delta_{\mathrm{history}}$ says that the visible coordinates must
belong to the same decision maker. A positive $\Delta_{\mathrm{domain}}$ says
that other domains contain information about the hidden domain. A positive
$\Delta_{\mathrm{value}}$ says that one valuation rule fitted outside the
target domain transports that information. None of these inequalities follows
from using a common software package.

\begin{table}[H]
\centering
\caption{Evidence that links the domains}
\label{tab:identification-audit}
\footnotesize
\renewcommand{\arraystretch}{1.12}
\begin{tabularx}{\textwidth}{@{}>{\raggedright\arraybackslash}p{0.22\textwidth}
  >{\raggedright\arraybackslash}p{0.47\textwidth}X@{}}
\toprule
Question & Test and evidence & Interpretation \\
\midrule
Does a coherent decision maker matter? & In Table~\ref{tab:heldout-point}, independently shuffling the visible choices raises normalized mean absolute error from 0.227 to 0.288 and lowers within-two accuracy by 18.0 points & Task marginals alone do not generate the forecast \\

Does information cross domain boundaries? & The held-out whole-domain forecast in Table~\ref{tab:heldout-point} has normalized mean absolute error 0.227, compared with 0.273 for the training-sample median & Choices outside the target domain predict the same decision maker inside it \\

Can the valuation map transfer? & In Table~\ref{tab:embedding-utility}, common utility fitted without target-domain outcomes has normalized mean absolute error 0.149 and within-two accuracy 63.16 percent, versus 0.234 and 40.42 percent for the task median & One state-to-valuation map predicts every omitted domain \\

Can the predictor transfer without a local target label? & In Table~\ref{tab:zero-label}, normalized mean absolute error is 0.277 for TabFM, 0.281 for the median, and 0.276 for Extra Trees & Cross-domain signal survives; the foundation model ties a local nonlinear learner on the primary loss \\

Is the main result only a flexible-learner effect? & In Table~\ref{tab:all-ablations}, with identical features and 100 labels, within-two accuracy is 69.91 percent for TabFM, 67.73 for Extra Trees, and 64.40 for tuned CatBoost & Local nonlinear learning explains most, but not all, of the reconstruction performance \\

Is the joint density native to the checkpoint? & The released decoder is scalar; dependence in Table~\ref{tab:heldout-joint} comes from an external residual vector & The checkpoint supplies conditional locations, not a native joint probability law \\
\bottomrule
\end{tabularx}
\begin{minipage}{\textwidth}
\footnotesize
\textit{Notes:} Mean absolute error is normalized by the feasible option-A
share span before averaging within decision maker. The common-utility exercise
uses task-masked frozen states and retains the 100 exact-target context labels;
only estimation of the valuation coefficients excludes target-domain
outcomes. The zero-label exercise removes target-domain labels and features and
is exploratory. Extra Trees is a randomised tree ensemble fitted locally;
CatBoost is a locally tuned gradient-boosted tree model.
\end{minipage}
\end{table}

The model's unity therefore does not rest on the fact that one checkpoint is
called 26 times. It rests on the conjunction of same-decision-maker transfer,
whole-domain masking, and the common valuation restriction in
equation~\eqref{eq:common-model-audit}. The zero-label experiment shows that a
small amount of transfer remains when local supervision is removed, but does
not establish a foundation-model advantage on normalized error. Likewise, the
joint-residual exercise is a useful hybrid forecast, not evidence that the
released scalar head is itself a joint distribution.

\subsection{From raw screens to Econographics}

Let $Y_i=(Y_{i1},\ldots,Y_{i,26})$ be the vector of raw switch responses used
in the prediction experiment. For covered econographic $c$, let
$\psi_{c\ell}$ be the authors' public transformation
\citep{ChapmanEtAl2022Data} for replicate
$\ell\in\{1,2\}$ and define
\begin{equation}
 M_{ic\ell}=\psi_{c\ell}(Y_i).
 \label{eq:author-construct-map}
\end{equation}
The maps include changes of economic orientation and, for ambiguity and
compound-lottery aversion, subtraction of the matched known-risk urn measure.
The 26 screens produce two replicates for 12 of the 21 econographics. The
known-risk urn screens are auxiliary inputs to two constructs; they are not a
thirteenth econographic.

Write $\rho_w(U,V)$ for the correlation of $U$ and $V$ under the authors'
survey weights. For constructs $c$ and $c'$, the two-replicate obviously
related instrumental variables correlation \citep{GillenSnowbergYariv2019}
reproduced here is
\begin{equation}
 R_{cc'}=
 \frac{
 |\mathcal P_{cc'}|^{-1}
 \sum_{(\ell,m)\in\mathcal P_{cc'}}
 \rho_w(M_{c\ell},M_{c'm})
 }{
 \left[
 \rho_w(M_{c1},M_{c2})
 \rho_w(M_{c'1},M_{c'2})
 \right]^{1/2}}.
 \label{eq:oriv-audit}
\end{equation}
The set $\mathcal P_{cc'}$ contains all four cross-replicate pairs in the
standard case. For ambiguity versus compound-lottery aversion, the replicated
analysis uses only opposite-replicate pairs because the two measures share the
same known-risk baseline and hence have nonorthogonal measurement errors. This
is the correction in the authors' code rather than a new measurement choice.

\begin{table}[H]
\centering
\caption{Two empirical objects in the paper}
\label{tab:measurement-audit}
\small
\renewcommand{\arraystretch}{1.13}
\begin{tabularx}{\textwidth}{@{}>{\raggedright\arraybackslash}p{0.20\textwidth}
  >{\raggedright\arraybackslash}p{0.34\textwidth}X@{}}
\toprule
Feature & Individual prediction experiment & Econographics reconstruction \\
\midrule
Outcome & 26 raw switch responses & 12 author-defined measures generated by equation~\eqref{eq:author-construct-map} \\
Coverage & Eight analysis domains in the multiple price list subset & 12 of the authors' 21 econographics; social-preference and overconfidence measures are outside the subset \\
Estimand & Prediction error for an equally weighted decision maker & Survey-weighted dependence among constructed measures \\
Measurement error & Raw response prediction; no disattenuation & Two-replicate correction in equation~\eqref{eq:oriv-audit} \\
Comparison scale & Raw row error and normalized option-A share error & Alignment of the 66 off-diagonal entries of the weighted correlation matrix \\
Audit & Feasible response maps and participant-level cross-fitting & Exact recovery of public construct columns and switch-row transformations, with maximum numerical error zero \\
Principal result & Held-out domain normalized mean absolute error of 0.227 & Common-utility rank alignment 0.939; leave-domain-out utility alignment 0.935 \\
\bottomrule
\end{tabularx}
\begin{minipage}{\textwidth}
\footnotesize
\textit{Notes:} The Econographics reconstruction uses the 800 training
participants, participant-cross-fitted predictions, and the survey weights in
the public replication file. Rank alignment is the Spearman correlation
between the 66 upper-triangle entries of the model-implied and observed
obviously related instrumental variables matrices. Validation and test
outcomes are not loaded.
\end{minipage}
\end{table}

This separation resolves the measurement question. The primary prediction
claim concerns raw response plans and gives each decision maker equal weight.
The claim about the dependence documented by \citet{ChapmanEtAl2023} is made
only after applying their transformations, survey weights, replicate
correction, and special shared-baseline adjustment. It does not equate the
project's eight analysis domains with the authors' six-component solution, and
it does not extend the reconstruction to the nine econographics absent from
the multiple price list subset.

Taken together, Tables~\ref{tab:identification-audit} and
\ref{tab:measurement-audit} identify the paper's central object. A frozen
foundation representation carries information about the same decision maker
across domains; one estimated random-utility map preserves most of that
predictive content and reconstructs the weighted dependence among the covered
econographics. A native joint head and utility over new primitive menus are
stronger objects, but neither is required for the reported common-index result.

\section{Additional Prediction Results}
\label{app:additional-results}

\subsection{Prediction without local labels}

The main design supplies 100 labels for the exact target. Table
\ref{tab:zero-label} reports a stricter validation exercise. For a task
holdout, no response to that task appears as a context label or feature. For a
domain holdout, every response in the domain is removed and context labels come
only from the other seven domains. Task and domain names are excluded; shared
numeric columns encode payoffs, probabilities, delays, endowments, social
payoffs, frames, and feasible rows.

\begin{table}[H]
\centering
\caption{Exploratory prediction without local labels}
\label{tab:zero-label}
\small
\setlength{\tabcolsep}{7pt}
\begin{tabular}{lrrrr}
\toprule
& \multicolumn{2}{c}{Leave one task out} & \multicolumn{2}{c}{Leave one domain out} \\
\cmidrule(lr){2-3}\cmidrule(lr){4-5}
Predictor & Normalized MAE & Within 2 (\%) & Normalized MAE & Within 2 (\%) \\
\midrule
Frozen TabFM       & 0.260 & 36.2 & 0.277 & 32.4 \\
Extra Trees        & 0.255 & 38.0 & 0.276 & 29.9 \\
Ridge              & 0.295 & 34.1 & 0.306 & 31.8 \\
Pooled median      & 0.279 & 30.5 & 0.281 & 28.1 \\
\bottomrule
\end{tabular}
\begin{minipage}{0.95\textwidth}
\footnotesize
\textit{Notes:} The task design contains 2,600 predictions for 100 validation
participants. The outcome-blind domain assignment contains 324. No test
outcome is loaded. These exercises were designed after the original held-out
test and are exploratory.
\end{minipage}
\end{table}

TabFM improves on pooled and linear forecasts without an exact-target label.
Extra Trees is better by 0.005 normalized MAE in the task design. In the domain
design, the two nonlinear predictors are statistically tied. The experiment
therefore establishes cross-task transfer but not a foundation-model advantage
unique to pretraining.

The TabFM wrapper removes context-constant columns and clips query values to
the observed context range. When a whole mechanism is absent, its indicator may
be removed and its payoff values may be clipped. The zero-label design is thus
also an extrapolation test of the public preprocessing pipeline. Theory-
generated contexts can cover target mechanisms without using a human target
label, which motivates Appendix~\ref{app:structural-adaptation}.

\subsection{External replication in a broader behavioural battery}
\label{app:sro-replication}

The Self-Regulation Ontology provides a substantially different test. Its
public first-wave release contains 522 participants, 37 behavioural tasks, and
129 task-level dependent variables spanning cognitive control, working memory,
learning, reasoning, risk, stopping, and intertemporal choice
\citep{EisenbergEtAl2019}. For every target task, I remove all of its dependent
variables and predict them from the same decision maker's other 36 tasks. Each
exact target variable receives 100 opposite-fold labels. Missing values are
preserved, and preprocessing and tuning use only those context rows.

\begin{table}[H]
\centering
\caption{External hidden-task prediction in the Self-Regulation Ontology}
\label{tab:sro-external}
\small
\setlength{\tabcolsep}{5.0pt}
\begin{tabularx}{\textwidth}{@{}>{\raggedright\arraybackslash}Xrrr@{}}
\toprule
Learner and information & Normalized MAE & Within 0.5 SD & Within 1 SD \\
\midrule
Frozen TabFM, coherent decision-maker behaviour & \textbf{0.7521} & \textbf{0.4148} & \textbf{0.7169} \\
Tuned CatBoost, coherent decision-maker behaviour & 0.7647 & 0.4072 & 0.7107 \\
Ridge, coherent decision-maker behaviour & 0.7732 & 0.3991 & 0.7031 \\
Factor ridge, coherent decision-maker behaviour & 0.7779 & 0.3971 & 0.7002 \\
Population median & 0.8281 & 0.3758 & 0.6684 \\
Frozen TabFM, profile only & 0.8322 & 0.3661 & 0.6638 \\
Frozen TabFM, shuffled context labels & 0.8450 & 0.3596 & 0.6540 \\
Frozen TabFM, shuffled decision maker & 0.8511 & 0.3559 & 0.6510 \\
\bottomrule
\end{tabularx}
\begin{minipage}{0.95\textwidth}
\footnotesize
\textit{Notes:} The analysis contains 522 participants, 37 task blocks, 129
target variables, and 18,442 nonmissing participant-task units. Absolute error
is divided by the exact context-label standard deviation and then averaged
within participant and task. The final columns report the share of outcomes
within one half or one context standard deviation. The decision-maker shuffle
independently reassigns every visible task block across participants. All TabFM
weights remain frozen.
\end{minipage}
\end{table}

The participant-clustered reduction in normalized MAE relative to tuned
CatBoost is 0.0127, with 95 percent interval $[0.0100,0.0154]$; TabFM is better
for 66.7 percent of participants. The reduction is 0.0767 relative to the
population median, interval $[0.0698,0.0835]$, and 0.0996 relative to the
decision-maker-shuffle forecast, interval $[0.0926,0.1067]$. TabFM beats CatBoost in
nine of ten prespecified behavioural families. Thus the same ordering appears
outside the economic multiple-price-list battery: a nonlinear learner captures
most of the predictable structure, the frozen foundation model retains a
modest advantage, and destroying decision-maker identity produces the largest loss.

\subsection{Joint forecasts with common marginals}

Two-fold cross-fitting predicts all 26 tasks while hiding the target domain.
Within each domain, normalized actual-minus-predicted vectors form an empirical
residual pool. A joint forecast adds a whole sampled residual vector to the
TabFM conditional mean. An independent forecast samples every coordinate from
the same residual matrix. Balanced permutations give each coordinate the same
multiset of draws, so the two forecasts have exactly the same marginal
predictive distributions.

\begin{table}[H]
\centering
\caption{Held-out joint forecast of a randomly hidden domain}
\label{tab:heldout-joint}
\footnotesize
\setlength{\tabcolsep}{3.0pt}
\begin{tabular}{lcccc}
\toprule
Forecast & Energy & Variogram & Marginal CRPS & Coverage 50/80/95 \\
\midrule
TabFM mean + joint residual vector & 0.3170 & 0.0617 & 0.1566 & 49/77/90 \\
TabFM mean + independent residuals & 0.3246 & 0.0903 & 0.1566 & 51/76/88 \\
Population empirical joint         & 0.3536 & 0.0614 & 0.1793 & 55/82/93 \\
Population empirical independent   & 0.3653 & 0.1127 & 0.1793 & 57/75/87 \\
\bottomrule
\end{tabular}
\begin{minipage}{0.95\textwidth}
\footnotesize
\textit{Notes:} There are 100 held-out participants and 324 assigned hidden
outcomes. Lower is better for all three proper scores. CRPS denotes continuous
ranked probability score. Coverage is percent. Joint and independent versions
have equal coordinate-wise marginal forecasts by construction. Variogram
power is 0.5.
\end{minipage}
\end{table}

Joint residual sampling reduces the energy score by 0.00752, with 95 percent
interval $[0.00412,0.01079]$, and the variogram score by 0.02860, interval
$[0.01650,0.04024]$. The improvement occurs in all eight domains. Relative to
the population empirical joint, the hybrid reduces energy by 0.03660 and mean
marginal CRPS by 0.02273. Its variogram score is essentially the same. TabFM
adds decision-maker-specific marginal location; the empirical residual vector adds
dispersion and dependence.

In the full-sample analysis, joint TabFM residual sampling has energy 0.30883
versus 0.31550 under independent sampling and variogram 0.06141 versus 0.08958.
Mean marginal CRPS is identical to numerical precision. Relative to the
population empirical joint, the hybrid improves energy and CRPS but has a
slightly worse variogram score. The scalar checkpoint therefore supports a
strong hybrid distribution, not a native TabFM joint density.

\subsection{Whole-domain results by interpretation}

The raw full-sample domain mask improves average MAE in seven of eight domains
and within-two accuracy in six. Time discounting is the principal exception:
the empirical distribution is concentrated, so the population median is
difficult to beat. Distributional preferences also favour the median under the
within-two loss while weakly favouring TabFM under MAE. These differences are
why the paper reports several losses and avoids a blanket dominance claim.

The validation-shrunken sensitivity reaches 47.55 percent within two rows but
has MAE 3.480, compared with 45.81 percent and 3.455 for raw TabFM. The matched
population median reaches 40.26 percent and 4.056. Shrinkage trades large-error
improvement for exact local concentration. It was locked for the confirmatory
test but is not reselected on the complete sample.

\section{Additional Geometry and Measurement Details}

Section~\ref{sec:geometry-main} reports the central geometric evidence. This
appendix documents the extracted states, projection, exploratory aggregation,
ablation geometry, probes, and forward-pass audit.

\subsection{Decision-maker and task states and linear CKA}

For every query defined by a decision maker and a task, I expose three states from
the frozen network.
The pre-ICL state is the 2,048-dimensional row-interaction output before labelled
examples are integrated. The contextual state is the 2,048-dimensional output
of the in-context transformer and is the primary representation. The
4,096-dimensional decoder hidden activation is the state immediately before
the scalar prediction. The complete extraction stores all three states for
$1{,}000\times26=26{,}000$ query cells. The exposed decoder path has exact
numerical parity with the packaged prediction method, and no array contains a
missing value.

Equation~\eqref{eq:cka} compares the participant-state matrices for two masked
tasks \citep{KornblithEtAl2019}. Before that comparison, one global principal
component analysis (PCA)
maps the 26,000 contextual states
to 64 dimensions and retains 75.9 percent of their variance. The mean state is
then removed separately by task and query fold. Fold centring prevents the two
different context samples from creating a common offset. CKA is invariant to
orthogonal rotation and isotropic rescaling and asks whether the same participants
occupy similar relative positions under two masked targets.

\subsection{A stable economic organization}

Figure~\ref{fig:task-nearness} contains three related objects. The observed
matrix reproduces the clustering problem that motivates Econographics. The
predicted matrix asks whether a model that forecasts each coordinate also
reproduces cross-task correlation. The contextual matrix asks whether the
internal decision-maker geometry induced by one masked task is near the geometry
induced by another.

The answer is sharply organized relative to the analysis-defined holdout
groups. Mean contextual CKA is 0.487 for two tasks in the same group and 0.206
across groups. Their difference, 0.281, has one-sided $p<0.001$ under 5,000
task-label permutations. Seven clusters maximize silhouette. Relative to the
eight analysis-defined groups, the partition has
adjusted Rand index 0.461 and normalized mutual information 0.807. The cluster
partition is identical across participant folds up to label permutation, and
the upper triangles of the two fold-specific CKA matrices have Spearman
correlation 0.794.

The embedding map is not merely organized by labels. Its upper triangle has
Spearman correlation 0.769 with the absolute observed response-correlation
matrix and 0.828 with the absolute predicted-correlation matrix; both
quadratic assignment procedure (QAP) tests give $p<0.001$. The observed and
predicted correlation matrices are even
closer, with rank correlation 0.945. Thus the model does not only reduce
coordinate-wise error. Across decision makers it reconstructs much of the
dependence pattern among tasks.

\subsection{The Chapman dependence matrix under common utility}

The raw-task comparison above is descriptive. The sharper exercise uses the
authors' economic constructs and measurement correction. I map observed and
cross-fitted predicted switch rows through the exact public transformations for
the 12 constructs covered by the 26 screens. I then reconstruct the
survey-weighted two-replicate ORIV correlation matrix, including the authors'
special cross-replicate correction for ambiguity and compound-lottery
aversion. The transforms recover both the public construct columns and the
constructs regenerated from actual switch rows with zero numerical error.

\begin{table}[H]
\centering
\caption{Common utility and the Chapman dependence structure}
\label{tab:chapman-utility-matrix}
\small
\setlength{\tabcolsep}{4.2pt}
\begin{tabular}{lrrrrr}
\toprule
Implied ORIV matrix & Pearson & Spearman & RMSE & Sign & Subspace \\
\midrule
Common utility & 0.959 & 0.939 & 0.142 & 0.864 & 0.965 \\
Utility fitted outside target domain & 0.956 & 0.935 & 0.136 & 0.879 & 0.936 \\
TabFM nonlinear decoder & 0.928 & 0.870 & 0.162 & 0.833 & 0.945 \\
Observed train sample versus full sample & 0.993 & 0.976 & 0.028 & 0.864 & 0.991 \\
\bottomrule
\end{tabular}
\begin{minipage}{0.96\textwidth}
\footnotesize
\textit{Notes:} Entries compare the 66 off-diagonal elements of the indicated
matrix with the observed train-sample ORIV matrix, except the final row, which
compares the 800-participant train matrix with the authors' full 1,000-person
matrix. Subspace is the overlap between the leading three eigenspaces. RMSE
denotes root mean squared error; ORIV denotes obviously related instrumental
variables.
Participant bootstrap 95 percent intervals for Spearman alignment are
$[0.837,0.963]$, $[0.831,0.959]$, and $[0.781,0.931]$ for the first three rows.
Each quadratic-assignment test uses 9,999 permutations and gives $p=0.0001$.
Validation and test outcomes are not loaded.
\end{minipage}
\end{table}

The common utility recovers the ordering of the dependence matrix more closely
than the nonlinear decoder. It also overstates the strength of dependence: its
mean absolute off-diagonal correlation is 0.266, compared with 0.167 in the
data. Equation~\eqref{eq:correlation-decomposition} separates this result into
systematic, residual, and cross covariances. The systematic covariance has
rank alignment 0.936 with the observed average-score covariance and 73.8
percent of its off-diagonal Frobenius norm. Residual covariance has 13.1
percent, while the two cross terms jointly have 28.6 percent. The systematic
utility therefore organizes most of the empirical pattern, but is not yet a
fully orthogonal conditional-mean decomposition.

This exercise uses task-masked states. The target response is absent, but
other tasks in the same domain may remain visible. Excluding the target domain
when fitting the valuation coefficients leaves rank alignment at 0.935, but
does not remove same-domain choices from the representation. The result
establishes a common predictive representation and a common valuation map; a
whole-domain-masked extraction is the stronger test of dependence generated
exclusively from other domains.

The resulting organization is descriptively economically recognizable. Time tasks and the two
distributional directions form separate tight blocks. The four probability-
weighting tasks form another. Mixed gain and loss tasks are near the loss only
tasks. Ambiguity, known risk, willingness to accept (WTA), and compound risk
occupy a broader uncertainty/valuation region. Willingness to pay (WTP) is
separated from WTA despite their common valuation label, echoing the WTA versus
WTP/inequality distinctions emphasized
in the original component analysis. Because the groups were defined for this
analysis and the interpretation follows inspection of the matrix, this is not
recovery of an independently estimated economic ontology.

\subsection{Exploratory aggregation of related raw task variants}

For a descriptive robustness exercise, I partition the 26 raw screens into 13
analysis-defined families of two related elicitation variants. I standardize
each raw switch response separately and average the two members before
correlation, parallel analysis, and Varimax rotation. This is not the
Econographics construction: the authors transform screens into economically
oriented measures, use the known-risk urn as an auxiliary baseline for
ambiguity and compound-lottery aversion, apply survey weights, and correct
correlations for measurement error using their repeated elicitations. The
aggregation here is neither ORIV nor a reliability correction.

Parallel analysis retains two components of this exploratory 13-family matrix,
explaining 44.6 percent of its standardized variation. At the unaggregated raw-
task level it retains seven components, explaining 68.7 percent. Neither count
is comparable to the six components in the authors' weighted, ORIV-corrected
PCA of 21 constructed econographics. I do not treat arithmetic averages of two
task-specific latent vectors as family embeddings: such averages are not
invariant to an otherwise innocuous task-specific rotation. The representation
claims therefore rest on the 26 task-level CKA objects above.

\subsection{Mechanical geometry and ablation evidence}

The most important caveat is that every task-masked query already contains the
other 25 choices. A raw vector with one coordinate missing will mechanically
inherit much of the decision-maker--task correlation structure. In the complete
sample, raw masked choices have rank alignment 0.905 with absolute observed
correlations. But their task similarity matrix is nearly degenerate. Mean CKA
within and between domains is close to one, and agreement across folds is only
0.117. High alignment alone is therefore
not evidence that the model discovered preferences.

The descriptive representation stress test uses matched projected states under
the information interventions. Valid decoder states have fold stability 0.848
and rank alignment 0.927 with observed behaviour. Choices only retains almost
all of that alignment, 0.922. It falls to 0.155 after shuffling decision-maker
identity, $-0.118$ after shuffling context labels, and 0.088 for profile only. A
five-choice history retains a weaker but substantial decoder geometry, with
alignment 0.690 and fold stability 0.486. The contextual state tells the same
story at an earlier model stage: observed-behaviour alignment is 0.679 under the
valid full input and 0.685 for choices only, but $-0.092$ and $-0.060$ under
the two destructive shuffles. These are all 1,000-participant estimates from the
same folds and global projection.

This ablation evidence is the bridge from data geometry to model geometry.
If a stable, domain-organized map remains after valid frozen inference but
disappears when the decision-maker vector or label map is broken, the representation
may be doing more than copying a correlation matrix. It is transforming the
available behavioural panel into a target-specific predictive organization.
That is evidence that the representation is responsive to coherent information
at the decision-maker level. It is not independent identification of a latent preference
structure and does not identify utility parameters. The zero-label experiment
in the next section tests transfer directly rather than inferring it from
geometry.

\subsection{Where the unified representation emerges}
\label{app:layerwise-lens}

The preceding analysis compares three exposed states. A locked train-only
replication follows the representation through all 24 blocks of the
in-context-learning (ICL) transformer. Work on TabPFN has used hidden-state
probes and an output-head lens for the same purpose
\citep{GuptaSethiKumar2026}. Because TabFM is a regression model, its analogue
of a language-model logit lens is a \emph{prediction lens}: after each block, I
apply the frozen final normalization and scalar prediction head. A raw lens can
be distorted when early and late layers use different coordinate systems. I
therefore pair it with a layer-specific ridge probe trained on one participant
fold and evaluated on the other, following the logic of the tuned lens
\citep{BelroseEtAl2023}.

Before the full extraction, I locked the 26 catalogue tasks, eight domain labels,
200 training participants, two folds of 100, one shared 256-dimensional random
projection, and a 99,999-draw task-label randomization test. Every target uses
all 100 participants in the opposite fold, so context identities are identical
across tasks. For each participant and target, the extractor saves the ICL
input and the state after every block. The manually exposed attention and
feed-forward updates reproduce the packaged TabFM forward pass exactly. No
validation or test record is loaded.

\begin{table}[H]
\centering
\caption{Layerwise decodability in frozen TabFM}
\label{tab:layerwise-probes}
\small
\setstretch{1.00}
\setlength{\tabcolsep}{7pt}
\begin{tabular}{lrrr}
\toprule
State & Final prediction & Observed response & Lens distance \\
\midrule
In-context-learning input & 0.843 & 0.636 & 0.325 \\
After block 15            & 0.953 & 0.641 & 0.153 \\
After block 24            & 0.950 & 0.639 & 0.000 \\
\bottomrule
\end{tabular}
\begin{minipage}{0.92\textwidth}
\footnotesize
\textit{Notes:} The first two columns are mean Pearson correlations across 26
task-specific ridge probes trained on one participant fold and tested on the
other. Lens distance is mean absolute distance from the final frozen output.
The exercise contains 5,200 train-only decision maker by task queries.
\end{minipage}
\end{table}

Table~\ref{tab:layerwise-probes} separates recovery of the
model's answer from recovery of behaviour. The raw prediction lens is
nonmonotone and first comes within 0.05 of the final output at block 22, which
confirms the value of the cross-fitted probe.

Figure~\ref{fig:layerwise-domain-cka} gives the main internal result. At the
ICL input, mean CKA is 0.681 within domains and 0.516 across domains, a
contrast of 0.165. At the output, the corresponding values are 0.537 and
0.301, a contrast of 0.236. The transformer lowers similarity broadly while
preserving substantially more common participant geometry within economic
domains. The locked increase in the within-minus-between contrast is 0.071.
None of 99,999 fixed-size domain-label permutations produces a gain as large,
so the plus-one one-sided randomization $p$-value is $10^{-5}$. The final contrast
is 0.227 in one participant fold and 0.206 in the other.

The organization is geometric rather than block modular. The 24-block update
signatures have cosine similarity above 0.9997 even across different domains.
TabFM does not reserve one collection of blocks for risk and another for time.
The same sequence of transformations carries domain-specific states and
increasingly distinguishes unrelated tasks while preserving more geometry
within domains. This is the internal organization expected from one unified
predictive model rather than a visible bundle of domain-specific estimators.
The task-mask representation still leaves other tasks from the target domain
visible. The whole-domain prediction design supplies the complementary
transfer test in which that local bridge is absent.

\section{Tabular Foundation Models for Economists}
\label{app:tabfm-economists}

This appendix explains the released TabFM architecture and its role in the
empirical design without presuming a machine-learning background. It separates
facts visible in the official model card and pinned source code from facts that
Google has not released. The description refers to the PyTorch regression
checkpoint used in this paper
\citep{KongDas2026,GoogleResearchTabFM2026,GoogleResearchTabFMRepo2026}.

\subsection{From an estimated model to an in-context predictor}

Let
\[
  \mathcal D_t=\{(x_j,y_{jt}):j\in\mathcal C_t\}
\]
be a context table for target $t$. Each row contains features $x_j$ and an
observed outcome $y_{jt}$. If $\Gamma$ is the conventional model's parameter
space, a supervised procedure estimates a task-specific vector
$\widehat\gamma_t(\mathcal D_t)\in\Gamma$ before it predicts a query row.
TabFM instead applies one set of pretrained parameters $\omega$ to the entire
prediction problem:
\begin{equation}
  \widehat y_{it}=f_{\omega}(x_i;\mathcal D_t).
  \label{eq:tabfm-operator}
\end{equation}
The context is data, not a parameter update. The public software uses the
familiar method name \texttt{fit}, but that call encodes columns, standardizes
the outcome, and stores the labelled context. It does not change $\omega$. The
prediction is therefore parameter-zero-shot but not label-zero-shot whenever
$\mathcal D_t$ contains outcomes from the target task.

The economic analogue is a decision rule with a pretrained prior over
statistical relationships. Instead of specifying a utility function and
estimating its parameters on $\mathcal D_t$, the model uses regularities learned
across many earlier synthetic prediction problems to infer a local conditional
mapping from the context. Equation~\eqref{eq:tabfm-operator} is a predictive
object. By itself it is not a utility representation, a causal model of the
decision maker, or a policy-invariant structural parameter.

\subsection{Architecture of the released checkpoint}

The input is a two-dimensional table. Rows are participants or observations; columns
are features. Ordinary language transformers process a one-dimensional word
sequence. TabFM instead alternates operations across the two axes of the table,
then compresses each row before performing in-context learning (ICL). Tables
\ref{tab:tabfm-representation} and \ref{tab:tabfm-prediction} give the
computation and its economic role.

\begin{table}[H]
\centering
\caption{The TabFM pipeline: representing a table}
\label{tab:tabfm-representation}
\footnotesize
\setstretch{0.98}
\renewcommand{\arraystretch}{1.08}
\begin{tabularx}{\textwidth}{@{}>{\raggedright\arraybackslash}p{0.17\textwidth}
  >{\raggedright\arraybackslash}p{0.40\textwidth}
  >{\raggedright\arraybackslash}X@{}}
\toprule
Stage & Public architecture & Interpretation in this paper \\
\midrule
Preprocessing &
Categorical values are numerically encoded, missing values are imputed,
constant columns are removed, and features and the regression target are
scaled using the context. &
The apparent \texttt{fit} step prepares the table; it does not train the
network. A target descriptor that is constant within a routed episode may be
removed here. \\
\addlinespace
Cells &
Each numerical or categorical cell is mapped through 32 learned Fourier
frequencies and a linear projection to 256 coordinates. Nearby feature slots
are grouped in threes. &
The raw coding of a switch response or payoff primitive becomes a learned
vector. This is a flexible numerical representation, not an economic utility
transformation. \\
\addlinespace
Columns &
Three induced set-attention blocks compare the same feature across context
rows. Each block has four attention heads and 256 learned inducing points. &
For a given feature, the model can compare the query participant with labelled
participants in the context without fitting a new coefficient. \\
\addlinespace
Rows &
Three row-attention blocks combine features within each participant. Eight learned
classification-summary (CLS) tokens collect the row information. Rotary
position embedding (RoPE) supplies feature-position information. The column
and row stages are then repeated. &
The model forms interactions among a participant's profile, visible choices, and
target descriptors. The eight summary tokens replace the full row by a compact
participant--target representation. \\
\bottomrule
\end{tabularx}
\begin{minipage}{\textwidth}
\footnotesize
\textit{Notes:} Dimensions describe TabFM 1.0.0. An attention head is one
parallel similarity calculation inside an attention block. Induced attention
uses a fixed set of learned intermediate points so that comparisons across many
rows are less costly. Fourier features are learned sine and cosine
transformations of a number. RoPE rotates internal coordinates as a function
of position.
\end{minipage}
\end{table}

\begin{table}[H]
\centering
\caption{The TabFM pipeline: prediction from labelled context}
\label{tab:tabfm-prediction}
\footnotesize
\setstretch{0.98}
\renewcommand{\arraystretch}{1.08}
\begin{tabularx}{\textwidth}{@{}>{\raggedright\arraybackslash}p{0.17\textwidth}
  >{\raggedright\arraybackslash}p{0.40\textwidth}
  >{\raggedright\arraybackslash}X@{}}
\toprule
Stage & Public architecture & Interpretation in this paper \\
\midrule
Compression &
The eight 256-coordinate summary tokens are concatenated into one
2,048-coordinate row vector. &
This is the pre-ICL representation extracted in the geometry analysis. \\
\addlinespace
Context &
A 24-block transformer with eight attention heads operates across compressed
rows. Observed outcomes are embedded only for context rows. The attention mask
prevents an unobserved query outcome from entering its prediction. &
The network converts labelled examples from other participants into a prediction rule
for the query participant. No Econographics gradient update occurs in a frozen run.
\\
\addlinespace
Output &
The regression decoder is a multilayer perceptron (MLP) with one scalar output.
The public classification checkpoint instead returns at most ten class logits.
&
The regression output is mapped to the nearest feasible multiple price list
switch row. The released head does not produce a joint distribution over all
tasks in a hidden domain. \\
\bottomrule
\end{tabularx}
\begin{minipage}{\textwidth}
\footnotesize
\textit{Notes:} An MLP is a short stack of linear maps and nonlinearities.
Observed outcomes enter only through labelled context rows; query outcomes are
never supplied to the model.
\end{minipage}
\end{table}

The checkpoint uses a feed-forward expansion factor of four and the
Swish-Gated Linear Unit (SwiGLU) activation. SwiGLU is a gated nonlinear
transformation inside each transformer
block. These details affect capacity and optimization, but the economically
important division is simpler. The first stages construct a representation of
each row from the two axes of the table. The 24-block ICL transformer then uses
labelled rows to turn that representation into a target-specific conditional
prediction.

\subsection{Why the model is called a foundation model}

TabFM's parameters were learned before the Econographics table was supplied.
Google reports training on hundreds of millions of tables drawn dynamically
from structural causal models (SCMs). In schematic form, pretraining solves
\begin{equation}
  \omega^\ast \in \arg\min_{\omega}
  \mathbb E_{\tau\sim P_{\mathrm{pre}}}\left[
    \sum_{i\in\mathcal Q_\tau}
    \ell\bigl(y_i,
    f_{\omega}(x_i;\mathcal D_\tau)\bigr)
  \right],
  \label{eq:tabfm-pretraining}
\end{equation}
where $\tau$ indexes a synthetic table, $\mathcal D_\tau$ is its labelled
context, $\mathcal Q_\tau$ is its query set, and $P_{\mathrm{pre}}$ is the
synthetic pretraining distribution. The function
$\ell:\mathbb R\times\mathbb R\rightarrow\mathbb R_+$ is the developer's
per-query pretraining loss. It is left unspecified because the training code
and exact loss have not been released; equation~\eqref{eq:tabfm-pretraining} is
a schematic definition of the training objective, not a claim about its
undisclosed implementation. The same $\omega^\ast$ can then be reused
on a new table. That reuse across data-generating processes is the
foundation-model claim.

Equation~\eqref{eq:tabfm-pretraining} also locates the main uncertainty. Google
has released the weights, inference source, model card, and benchmark results,
but not $P_{\mathrm{pre}}$, the synthetic generator, sampled pretraining tables,
the pretraining code, or a technical report. We therefore know the broad class
of prior---random functions connected through SCMs---but not its exact support,
mixture weights, or overlap with the statistical form of this experiment. The
released materials report training on synthetic tables; the unavailable
generator and sampled corpus prevent a more detailed account of that prior.

\begin{table}[H]
\centering
\caption{What is known about TabFM 1.0.0}
\label{tab:tabfm-disclosure}
\small
\renewcommand{\arraystretch}{1.15}
\begin{tabularx}{\textwidth}{@{}>{\raggedright\arraybackslash}p{0.27\textwidth}
  >{\raggedright\arraybackslash}p{0.29\textwidth}
  >{\raggedright\arraybackslash}X@{}}
\toprule
Object & Public status & Consequence for interpretation \\
\midrule
Architecture and inference & Source and checkpoint configuration are public &
The forward computation and dimensions in
Tables~\ref{tab:tabfm-representation} and \ref{tab:tabfm-prediction}
can be inspected and pinned. \\
Pretrained parameters & Public under a noncommercial weight licence &
The exact checkpoint used here can be hashed and archived for academic
reproduction. \\
Broad pretraining description & Hundreds of millions of synthetic SCM tables &
The model carries a broad tabular prior, but its economic content is not
identified by this description. \\
Generator and sampled corpus & Not public &
The released materials do not establish the exact training-data composition
or overlap with the present battery. \\
Technical report and training code & Not public as of August 2026 &
Claims about training loss, sampling weights, and ablations must not be filled
in by conjecture. \\
\bottomrule
\end{tabularx}
\end{table}

\subsection{The exact role of TabFM in the experiments}

In the target-routed experiments, one context table contains 100 labelled
decision makers for the exact target. The query row contains the focal decision maker's
profile, visible choices, and the target feature schema, but not the target
outcome. The regression target is the share of feasible rows on which option A
is chosen. The scalar forecast is projected to the nearest feasible switch
code. All primary weights remain frozen. Thus the experiment tests whether one
pretrained prediction operator can repeatedly use a coherent decision maker's visible
behaviour across targets. It does not test prediction without local labels.

The zero-label experiments change the information set, not the meaning of
``frozen.'' They withhold every label from the target task or domain while
retaining labels from other targets. This is the sharper test of cross-task
transfer. It is also more exposed to preprocessing: a mechanism indicator or
payoff feature that never varies in the remaining context can be removed, and a
query value outside the context range can be clipped. These are architectural
constraints, not evidence that cross-domain prediction is impossible.

The joint forecast in the main text adds a vector sampled from the
training-only empirical residual distribution to 26 scalar TabFM means. That
construction is useful because it
holds marginal predictive distributions fixed while testing dependence. It is
nonetheless a hybrid. Dependence comes from the external residual distribution,
not from the released scalar decoder. A native joint claim requires a common
probabilistic head that assigns probability to feasible switch-row vectors.

\subsection{Structural adaptation used in the extension}

The structural extension preserves the pretrained backbone and attaches
low-rank adaptation (LoRA) matrices to the two linear layers of the scalar
decoder. LoRA writes a decoder map as
\[
  W^\prime = W + \frac{\alpha}{r}BA,
\]
where $W$ is frozen, $A$ and $B$ are trained, $r=8$ is the adapter rank, and
$\alpha=16$ fixes the scale. The implemented decoder adapter has 81,928
trainable parameters; the rest of the checkpoint is frozen. Training
alternates a randomly hidden task and a randomly hidden domain and minimizes a
smooth absolute-error loss on the context-standardized share of A choices.

This adapter changes only the decoder. It can reweight information already
present in TabFM's contextual state but cannot change how the backbone
constructs that state. The reported gain therefore identifies a more useful
mapping from the frozen state, not a theory-induced change in the foundation
representation.

\subsection{Reading the evidence correctly}

Four distinctions summarize the interpretation.

\begin{enumerate}
  \item Frozen weights do not mean no supervision. Exact-target context labels
  provide local supervision unless the zero-label design removes them.
  \item One common checkpoint does not by itself imply one common utility
  function. The established object is a shared conditional prediction
  operator.
  \item Accurate scalar predictions do not imply a joint probability model.
  Marginal calibration and within-decision-maker dependence require a probabilistic
  head and proper multivariate scores.
  \item At equal compute, joint and independent-marginal adapters are tied.
  The gain belongs to marginal theory, not cross-domain linkage.
\end{enumerate}

\clearpage
\section{Literature and Scope}
\label{app:literature}

This appendix places the paper within four literatures. The organizing
distinction is between the representation of a decision problem, the valuation
of that representation, and the source of stochastic choice. The paper changes
the first object, imposes a common restriction on the second, and uses a
standard shock specification for the third.

\begin{table}[H]
\centering
\caption{Where the paper enters the literature}
\label{tab:literature-map}
\small
\renewcommand{\arraystretch}{1.12}
\begin{tabularx}{\textwidth}{@{}>{\raggedright\arraybackslash}p{0.20\textwidth}
  >{\raggedright\arraybackslash}p{0.27\textwidth}
  >{\raggedright\arraybackslash}p{0.23\textwidth}
  >{\raggedright\arraybackslash}X@{}}
\toprule
Literature & Representative work & Principal object & This paper \\
\midrule
Random utility and stochastic choice
& \citet{GulPesendorfer2006}, \citet{FudenbergIijimaStrzalecki2015},
  \citet{KitamuraStoye2018}
& Restrictions on stochastic choice generated by utility maximization
& Tests whether one valuation rule operates after unlike decision problems are
  mapped into common coordinates \\

Ordered random utility
& \citet{ApesteguiaBallesterLu2017},
  \citet{ApesteguiaBallester2025}
& Ordered types, ordered choices, and the empirical content of the map from
  types to utilities
& Uses a high dimensional frozen representation and restricts its map to a
  common scalar valuation index across domains \\

Bounded rationality and noisy choice
& \citet{CaplinDean2015}, \citet{CattaneoEtAl2020},
  \citet{BarseghyanMolinariThirkettle2021}
& Information, consideration, cognition, or noise within a specified choice
  environment
& Does not select among mechanisms; asks whether their joint behavioural residue
  is portable across environments \\

Preference measurement and machine learning
& \citet{BarskyEtAl1997}, \citet{ChoiEtAl2014},
  \citet{FalkEtAl2018}, \citet{FudenbergLiang2019}
& Stable preference measures, decision quality, and prediction as a route to
  model improvement
& Moves from association among measures to hidden domain prediction and then
  compresses the predictor into random utility \\
\bottomrule
\end{tabularx}
\begin{minipage}{\textwidth}
\footnotesize
\textit{Notes:} The table selects papers that define the closest economic
objects. It is not intended as a catalogue of domain specific models. ``Common''
refers to the valuation rule and decoder, not to identical representations or
identical predicted choices across domains.
\end{minipage}
\end{table}

\subsection{Random utility and its empirical content}

Random utility gives stochastic choice economic content by requiring observed
choice probabilities to arise from maximization. Choice based foundations
state the observable restrictions implied by particular versions of that
idea. \citet{GulPesendorfer2006} characterize random expected utility over
lotteries. \citet{FudenbergIijimaStrzalecki2015} characterize perturbed utility
representations. \citet{KitamuraStoye2018} develop an implementable
nonparametric test of random utility for heterogeneous demand, and
\citet{KonoSaitoSandroni2026} show how unobservable alternatives change the
rationalizability restrictions. Dynamic choice adds restrictions on the
persistence of tastes and learning that are absent from static marginals
\citep{FrickIijimaStrzalecki2019}.

The closest theoretical comparison is ordered random utility.
\citet{ApesteguiaBallesterLu2017} characterize random utility over a
single crossing family. \citet{ApesteguiaBallester2025} distinguish two model
components: a map from ordered latent types to utilities and a distribution
over those types. Their central result for the present paper is that fixing the
type distribution alone has no empirical cost when the type to utility map
remains unrestricted. Restrictions on that map carry empirical content, and
jointly restricting the map and the distribution adds further content.

The hybrid model has the same conceptual decomposition but a different data
object. The frozen transformer creates a decision maker by task state. The
common coefficient vector maps that state into scalar valuation. The response
plan utility and extreme value disturbances turn valuation into choice
probabilities. Thus the learned state is not itself the economic conclusion.
The cross domain restriction is imposed by using the same state to valuation
map for all domains and, more sharply, by predicting a domain whose outcomes
did not fit that map. The current experiment observes discrete response plans
from a panel of decision makers, not continuous cumulative choice
distributions. The results therefore do not establish the axioms in
\citet{ApesteguiaBallester2025}. They use that paper's decomposition to locate
the restriction that makes the learned model economically testable.

\subsection{Bounded rationality, attention, and choice noise}

Several top journal literatures give stochastic choice mechanisms other than
random taste shocks. Costly information can make apparent mistakes optimal
\citep{CaplinDean2015}. Random attention separates preferences from the set of
alternatives considered \citep{CattaneoEtAl2020}. Limited consideration can be
estimated jointly with risk preferences in rich discrete choice environments
\citep{BarseghyanMolinariThirkettle2021}. Aggregate choices can identify
heterogeneity in cognitive procedures under suitable structure
\citep{DardanoniEtAl2020}. Deliberate randomization provides another
behavioural interpretation and can violate the regularity property associated
with random utility \citep{CerreiaVioglioEtAl2019}. Response times can separate
preference from noise under conditions that choice data alone cannot provide
\citep{AlosFerrerFehrNetzer2021}.

This paper does not use predictive success to choose among these mechanisms.
A common representation may encode attention, cognition, framing, or
preference. The individual shuffle and profile ablations establish that the
useful signal belongs to a coherent decision maker's behavioural history. They
do not by themselves identify why a particular response was stochastic. The
economic gain is different: once the state contains that history, one common
valuation rule transports it across risk, time, loss, valuation, ambiguity,
and distributional problems.

\subsection{Preference heterogeneity across domains}

Economists have long measured several preference dimensions in the same
people. \citet{BarskyEtAl1997} connect experimentally elicited preference
parameters and behavioural heterogeneity in a representative survey.
\citet{ChoiEtAl2014} show that consistency with utility maximization varies
across people and predicts wealth. \citet{FalkEtAl2018} measure risk, time,
and social preferences across countries and document both within and between
country heterogeneity. Measurement error is central when persistent
individual traits are recovered from repeated choices
\citep{GillenSnowbergYariv2019,AguiarKashaev2021}.

\citet{ChapmanEtAl2023} take the decisive next step for this paper. They study
many behavioural phenomena in the same representative experimental panel,
document common empirical structure, and conclude that no model they examine
explains the full collection. Their object is the joint measurement of
econographics. The object here is a conditional prediction: after one domain
is hidden, can the remaining choices recover that decision maker's responses,
and can one utility map do so for every domain? Prediction turns shared
variation into an out of sample restriction. The response plan random utility
model then asks how much of that predictive structure can be expressed by one
valuation direction.

\subsection{Machine learning as an input to economic theory}

The machine learning literature in economics distinguishes prediction from
structural interpretation \citep{MullainathanSpiess2017}. A particularly close
precedent is \citet{FudenbergLiang2019}, who use machine learning to identify
regularities missed by existing models and then improve the economic model.
\citet{Liang2026} develops this broader program of using machine learning to
generate, clarify, and improve economic models. The present paper follows that
logic in reverse order from a pretrained model. It first establishes portable
prediction, then asks whether a severe economic restriction can recover the
prediction.

Tabular foundation models make this exercise possible because their parameters
are learned before the local table is observed
\citep{HollmannEtAl2025,KongDas2026}. The main comparison is therefore not only
between predictive algorithms. A locally fitted nonlinear learner measures the
information available in the experimental table. The frozen checkpoint asks
whether a previously learned inference rule transports that information. The
hybrid asks whether the transported information can be valued by one random
utility model. These are respectively statements about local predictability,
portable representation, and common economic valuation.

\section{Structural Synthetic Adaptation}
\label{app:structural-adaptation}

\subsection{Primitive actions and formal verification}
\label{app:primitive-action}

The response plan is the observed economic alternative in the current
experiment, so equations \eqref{eq:plan-systematic-utility}--
\eqref{eq:plan-random-utility} are sufficient for the cross-domain test. A
stronger model assigns utility directly to the primitive actions within each
row. Let $Z_{tr}$ describe the probabilities, payoffs, dates, or allocations
in row $r$, and let $j\in\{A,B\}$ identify the action. An action encoder and
the same common valuation give
\[
 e_{itrj}=\Phi^{\mathrm{act}}_{\omega}
 (X_i,H_{i,-t},Z_{tr},j,\mathcal D_t),
 \qquad v_{itrj}=\alpha+\beta^\top\Pi e_{itrj}.
\]
With independent type I extreme value shocks,
\begin{equation}
 \Pr(Y_{itr}=B\mid X_i,H_{i,-t},Z_{tr},\mathcal D_t)
 =\Lambda\!\left(v_{itrB}-v_{itrA}\right).
 \label{eq:action-embedding-rum}
\end{equation}
This form makes counterfactual changes in prizes, probabilities, dates, and
allocations explicit. The current extraction provides decision maker by task
states rather than separate action embeddings, so the empirical results
identify utility over feasible response plans. Primitive policy comparisons
add the requirement that action representations remain stable as primitives
change.

The encode, value, and choose factorization is formally verified in Lean. The
proof verifies representation invariance and prediction factorization as
algebraic properties of the architecture. It does not establish identification,
estimation consistency, the disturbance distribution, or counterfactual
stability. The formal proof, pinned toolchain, and verification instructions
are included in the replication package; Appendix~\ref{app:reproducibility}
describes the reproduction protocol.

\subsection{Economic generators}

The structural extension builds synthetic participants from familiar marginal
models. Risk and loss screens use reference-dependent value curvature, loss
aversion, and probability weighting. Time screens use exponential and present-
biased discounting. Valuation screens allow reference-dependent gaps between
willingness to accept and willingness to pay. Distributional screens use
inequality-aversion parameters. Ambiguity and compound-lottery screens add
domain-specific sensitivity parameters. Lapse noise and heterogeneous
parameter draws convert deterministic switch points into response
distributions.

The first generator joins these parameters inside a synthetic decision maker and
alternates a randomly hidden task with a randomly hidden domain. It is a
working economic prior, not an estimated structural model of the human sample.
The matched generator independently permutes complete domain blocks across
synthetic identities. It therefore holds every marginal domain distribution,
within-domain choice vector, and task histogram fixed while removing only
cross-domain identity, as defined in equation~\eqref{eq:joint-prior-value}.

\subsection{Adapter and training protocol}

The TabFM backbone contains 1,647,782,989 frozen parameters. Low-rank
adaptation (LoRA) matrices are attached to the two linear layers of the scalar
decoder. For a decoder matrix $W$,
\[
 W'=W+\frac{\alpha}{r}BA,
\]
where rank $r=8$, scale $\alpha=16$, and only $A$ and $B$ are trained. The
adapter has 81,928 parameters, or 0.005 percent of the combined model. Training
uses AdamW with learning rate 0.001 and a smooth absolute-error loss on the
context-standardized share of A choices.

The run resumes the 1,200-step adapter and continues to 20,000 steps. Each step
uses 512 context rows and 64 supervised queries. Task-mask and whole-domain
synthetic validation are computed every 400 steps. The 12,400-step checkpoint
is selected because it minimizes synthetic whole-domain share MAE. No human
validation or test outcome enters checkpoint selection.

\begin{table}[H]
\centering
\caption{Synthetic learning curve}
\label{tab:synthetic-learning-curve}
\small
\begin{tabular}{rrrrr}
\toprule
Step & Task MAE & Task within 2 & Domain MAE & Domain within 2 \\
\midrule
0      & 0.19122 & 0.4596 & 0.19214 & 0.4750 \\
400    & 0.18947 & 0.4788 & 0.18446 & 0.5135 \\
1,200  & 0.18720 & 0.4827 & 0.18304 & 0.5327 \\
12,400 & 0.18238 & 0.5077 & 0.18022 & 0.5135 \\
19,600 & 0.18226 & 0.5058 & 0.18155 & 0.5269 \\
20,000 & 0.18364 & 0.4885 & 0.18178 & 0.5154 \\
Indep., 15,200 & 0.19017 & 0.5000 & 0.18735 & 0.4731 \\
\bottomrule
\end{tabular}
\begin{minipage}{0.92\textwidth}
\footnotesize
\textit{Notes:} MAE denotes mean absolute error on the normalized share scale.
Step 12,400 minimizes whole-domain MAE in the joint arm; step 19,600 minimizes
its task MAE. Step 15,200 minimizes whole-domain MAE in the equal-compute
independent-marginal arm. The holdout is synthetic and uses seed 20260917.
\end{minipage}
\end{table}

Most gains arrive by step 12,400 and then plateau. Relative to step zero, the
best checkpoints reduce synthetic task error by 4.68 percent and domain error
by 6.20 percent. Training duration was a modest constraint, but continuing the
same fixed synthetic prior is no longer the main bottleneck.

\subsection{Human validation}

The selected domain checkpoint is evaluated on the designated 100-participant
validation split. The design contains 324 hidden-domain forecasts, 1,040
synthetic context rows, and 80 features. A task-level median adjustment is
estimated only from the 800-participant human training split. This anchor corrects
synthetic-to-human marginal level differences without using validation or test
outcomes.

\begin{table}[H]
\centering
\caption{Exploratory human validation of structural adaptation}
\label{tab:structural-validation}
\small
\setlength{\tabcolsep}{7pt}
\begin{tabular}{lrrr}
\toprule
Method & Normalized MAE & Row MAE & Within 2 (\%) \\
\midrule
Human training-population median & 0.2216 & 3.535 & 46.25 \\
Independent, 15,200, anchored & 0.2334 & 3.715 & 43.75 \\
Joint theory, step 12,400, anchored & 0.2371 & 3.768 & 44.75 \\
Joint theory, step 1,200, anchored  & 0.2444 & 3.875 & 41.75 \\
Frozen TabFM, anchored              & 0.2572 & 4.073 & 41.00 \\
Frozen TabFM, unanchored            & 0.2918 & 4.588 & 26.50 \\
Joint theory, step 12,400, unanchored & 0.2942 & 4.603 & 26.25 \\
Independent, 15,200, unanchored & 0.2979 & 4.650 & 25.00 \\
\bottomrule
\end{tabular}
\begin{minipage}{0.95\textwidth}
\footnotesize
\textit{Notes:} The checkpoint is selected on synthetic whole-domain MAE.
Anchoring uses only human training-sample task medians. The sealed test split is
not loaded or scored.
\end{minipage}
\end{table}

The anchored step-12,400 adapter reduces normalized error by 0.00726 relative
to step 1,200, with participant-bootstrap interval $[0.00166,0.01296]$. It
reduces error by 0.02007 relative to anchored frozen TabFM, interval
$[0.01126,0.02877]$. Within-two accuracy rises by 3.0 and 3.75 percentage points,
respectively. The adapter remains worse than the human population median by
0.01552 normalized error, interval $[0.00377,0.02741]$ in favour of the median.
Unanchored transfer remains statistically tied with the early adapter and
frozen TabFM.

The equal-compute linkage contrast isolates the source of this gain. Before
anchoring, joint theory reduces normalized error relative to independent
margins by 0.00371, with participant-bootstrap interval
$[-0.00193,0.00937]$ and one-sided sign-flip $p=0.105$. After the identical
training-sample anchor is applied to both arms, the joint-minus-independent
reduction is $-0.00368$, interval $[-0.00921,0.00185]$, with $p=0.902$ for a
joint-theory improvement. The sign reversal and intervals spanning zero show
that the current synthetic cross-domain links add no stable human-validation
gain beyond the matched marginal theories.

The domain pattern is heterogeneous. Error falls most for time discounting,
distributional preferences, ambiguity, compound risk, and losses. It is almost
unchanged for valuation and known risk, and rises by 0.0205 for probability
weighting. These small domain cells diagnose misspecification rather than
identify separate treatment effects. They reinforce the interpretation in the
main text: standard marginal theories add useful restrictions, while the
current synthetic links among those theories add no detectable transfer.

\subsection{Computation and reproducibility checks}

The joint adapter was trained on one central processing unit node provided by the
Digital Research Alliance of Canada. The allocation used 192 cores, with 96
PyTorch threads and no graphics processing unit. Continuing the adapter from
step 1,200 to step 20,000 took 10 hours, 54 minutes, and 57 seconds. Peak
resident memory was 15.7 GiB, and the run consumed 2,095.84 core-hours.

The equal-compute independent-marginal continuation used the same 192-core
shape and 96 PyTorch threads. It completed in 10 hours, 41 minutes, and 56
seconds, reached 15.7 GiB peak resident memory, and consumed 2,054.19
core-hours. Its 53 declared artifacts and scheduler logs passed remote and
local hash verification before the paired human-validation analysis.

Checkpoint selection followed the outcome-blind rule in the preceding
subsection: step 12,400 minimizes mean absolute error when an entire synthetic
domain is hidden. All declared model outputs and training logs passed an
integrity audit before analysis. A separate replay then reproduced all 324
human-validation forecasts exactly. These computational checks establish that
the reported coherent-arm estimates can be regenerated from the same public
inputs, model revision, and analysis specification; they do not use or reveal
the held-out test outcomes.

\section{Data, Software, and Reproducibility}
\label{app:reproducibility}

The human data are public in the Econographics replication package
\citep{ChapmanEtAl2022Data}. The analysis fixes the variable map, participant
split, target support, and labelled context before prediction. The one-time
confirmatory test is kept separate from the complete-sample and validation
analyses described elsewhere in the paper.

The TabFM checkpoint and inference source are public under the developer's
licence \citep{GoogleResearchTabFM2026,GoogleResearchTabFMRepo2026}. The exact
pretraining generator, corpus, and training code were not public as of August
2026. Reproduction of the forward pass is therefore possible; independent
reproduction of pretraining is not. Appendix~\ref{app:tabfm-economists}
separates released architecture facts from undisclosed training details.

The action-level encode, score, and decode restriction is formally verified in
Lean. The replication package includes the formal proof, a pinned Lean
toolchain, and an executable theorem audit. The empirical replication package
records the public input versions, fixed splits, context construction,
checkpoint identifiers, fitted structural parameters, prediction files, and
the information needed to regenerate every table and figure. It also includes
the participant-level resampling procedures used for statistical inference.

A publication-ready replication archive is prepared. It uses package-relative
instructions rather than machine-specific paths, so the empirical results and
formal verification can be reproduced without access to the author's computing
environment. The archive will be released with the journal-submission version.

\end{document}